\documentclass{article}
\usepackage{epsfig,psfrag,verbatim,amssymb,amsfonts,amsthm}
\usepackage{graphicx}
\usepackage{subfigure,url,amsmath}
\usepackage{subfigure,amsmath}
\usepackage{multirow}
\usepackage{color}
\usepackage{natbib}
\usepackage{vmargin} 
\usepackage[symbol]{footmisc}

\usepackage{booktabs}
\usepackage{xcolor}
\usepackage{color}
\usepackage{colortbl}
\usepackage{makecell}

\setpapersize{USletter}
\setmarginsrb{1truein}{1truein}{1truein}{1truein}{0pt}{0pt}{0pt}{0pt}
\begin{document}
\
\begin{center}
  {\large \bf Selection of inverse gamma and half-t priors for hierarchical models: sensitivity and recommendations}\\[2ex]
{\normalsize{Zachary Brehm\footnote[1]{Joint first authors} (1),
  Aaron Wagner$\null^{* }$ (1),
Erik VonKaenel$\null^{* }$ (1),
David Burton$\null^{* }$ (1),
Samuel J. Weisenthal~(1),
Martin Cole (1),
Yiping Pang (1),
and Sally W. Thurston\footnote[2]{Corresponding author:  sally\_thurston@urmc.rochester.edu} (1)}}\\[1ex]
( (1) University of Rochester, Department of Biostatistics and Computational Biology, Rochester, NY) \\
\end{center}
\renewcommand{\thefootnote}{\fnsymbol{footnote}}

\begin{abstract}
While the importance of prior selection is well understood, establishing guidelines for selecting priors in hierarchical models  has remained an active, and sometimes contentious, area of Bayesian methodology research. Choices of hyperparameters for individual families of priors are often discussed in the literature, but rarely are different families of priors compared under similar models and  hyperparameters.  Using simulated data, we evaluate the performance of inverse gamma and half-$t$ priors for estimating the standard deviation  of random effects in three hierarchical models: the 8-schools model, a random intercepts longitudinal model, and a simple multiple outcomes model.  We compare the performance of the two prior families using a range of prior hyperparameters, some of which have been suggested in the literature, and others that allow for a direct comparison of pairs of half-$t$ and inverse-gamma priors.   Estimation of very small values of the random effect standard deviation led to convergence issues  especially for the half-$t$ priors.  For most settings, we found that the posterior distribution of the standard deviation had smaller bias under half-$t$ priors than under their inverse-gamma counterparts. Inverse gamma priors generally gave similar coverage but had smaller interval lengths than their half-$t$ prior counterparts.  Our  results for these two prior families  will inform prior specification for hierarchical models, allowing practitioners to better align their priors with their respective models and goals.

\end{abstract}





\section{Introduction}

In some Bayesian models, it may not be clear how to choose reasonable prior distributions for  model parameters.  Often, conjugate priors are chosen so that the posterior can be derived in closed form, or so that Gibbs sampling can be used to obtain samples from the posterior.   Regardless of whether a conjugate prior is selected, one must also  specify values for the prior hyperparameters.  In many cases, these values can be selected so the prior has little influence on the posterior.  Although this is not necessarily the goal of prior specification, it is often of interest to be able choose relatively uninformative hyperparameters as long as their choice gives a proper posterior distribution.

Hierarchical models are often used for models with random effects when it is appropriate to share information across different subjects or units, a situation called ``borrowing strength''.  An example of such a model is
\begin{eqnarray}
  \label{eq:8schools1}
  \bar{y}_j \mid \theta_j, \sigma^2_j  & \sim &  N(\theta_j, \sigma^2_j) , \ j=1, \cdots, J \\
  \label{eq:8schools2}
   \theta_j \mid \tau^2  & \sim & N(\mu, \tau^2)
 \end{eqnarray}
The model in (\ref{eq:8schools1}) and (\ref{eq:8schools2}) is used for the well-known eight schools example \citep{GelmanBDA:2013}, in which eight schools carried out randomized experiments to estimate the effect of short term coaching on subsequent Scholastic Aptitude Test-Verbal (SAT-V) scores.  In that situation individual students' SAT-V scores were regressed on whether or not they had coaching (yes/no), and their prior PSAT-Verbal and  PSAT-Math scores.  In (\ref{eq:8schools1}),  $\bar{y}_j$ is the estimated coaching effect in school $j$, estimated by the slope for coaching from the school-specific regression fit to the $n_j$ subjects in the $j$th school.  The within-school variance, $\sigma^2_j$, is $\sigma^2 / n_j$ where $\sigma^2$ is assumed to be known.  

The choice of priors and their associated hyperparameters for the between-unit variability, $\tau^2$, or the between-unit standard deviation, $\tau$, in a hierarchical model has been discussed in a number of papers.
A uniform prior on the shrinkage parameter, $S=\frac{\sigma^2}{\sigma^2 + \tau^2}$, was shown to give a minimax estimator for the $J$ means in similar models under certain conditions \citep{Strawderman:1971}.
Implementation of the uniform shrinkage prior \citep{Daniels:1999} leads to proper posterior distributions while providing desirable frequentist properties for testing and estimation.
However, the uniform distribution on the shrinkage weight implies that $\sigma^2$ and $\tau^2$ are of the same order of magnitude,
a strong assumption that is not feasible in all data situations.

Some prior choices can lead to improper posteriors, such as $\mbox{log}(\tau) \sim \mbox{Uniform}(-\infty, \infty)$ \citep{GelmanBDA:2013}.  The prior $\tau \sim \mbox{Uniform}(0,A)$ yields a proper posterior, but results are sensitive to $A$ \citep{GelmanBDA:2013}.  
The property of conditional conjugacy makes the inverse gamma  (IG) prior for $\tau^2$ an attractive choice, and IG($\epsilon, \epsilon$) with small $\epsilon$ is often used \citep{GelmanPrior:2006}.  However, the posterior can be sensitive to the choice of $\epsilon$, and as $\epsilon \rightarrow 0$ the posterior becomes improper. 
The use of an IG($\epsilon, \epsilon$) prior has become contentious in part due to its
presence in the examples for the manual in the widely used software, BUGS  \citep{spiegelhalter2003winbugs}, where $\epsilon=0.001$ is the parameter used in the examples in the BUGS manual. 

  In Bayesian inference for generalized linear mixed models (GLMM) the choice of IG($\epsilon, \epsilon$) with small $\epsilon$ can be shown to place the majority of the prior mass away from zero \citep{Fong:2010}, leading to a marginal prior for the random effects that is Student's \textit{t} with $2\epsilon$  degrees of freedom.  When $\epsilon$ is very small, the resulting $t$ prior distribution has heavy tails.  An alternative strategy is to choose a conservative prior that deliberately gives more weight to smaller values of $\tau^2$  \citep{Gustafson:2006}.  This choice may be particularly advantageous when the consequences of inferring substantial variability of the random effects are costly or otherwise undesirable.

\cite{GelmanPrior:2006} recommend a folded half-t($\nu, s$) prior for the standard deviation, $\tau$, with $\nu$ degrees of freedom and scale $s$.  In particular, they recommend $\nu=1$ (a half-Cauchy prior), and suggest that $s$ be chosen to be a bit larger than the expected standard deviation of the random effects.  \cite{Polson:2012} also suggest the half-Cauchy as a default prior for $\tau$.  These authors provide additional justification for this choice from a classical risk perspective, and show that this prior leads to excellent quadratic risk properties for high-dimensional sparse data.   More recently, in his blog \url{https://github.com/stan-dev/stan/wiki/Prior-Choice-Recommendations} (edited 4/17/20, accessed 7/12/20), Gelman suggests using a half-t (HT) prior for $\tau$ with $s$ not too large, such as $s=1$ and with $\nu=4$ degrees of freedom, or a half-normal prior with $\nu$ very large.

It is clear from the literature that the choice of ``default'' priors is not a settled issue.
Debates about prior choices for noniformative priors center around the meaning and
acceptance of noninformativity.  A number of  researchers acknowledge the impossibility of truly noninformative priors, but this is not without  controversy. 
Although the use of half-t (HT) or half-Cauchy priors for $\tau$ is becoming more widespread, a half-Cauchy prior has been criticized for poor behavior during MCMC sampling, due to it's heavy tails \citep{PiironenVehtari:2015}.  

While one school of thought suggests using default hyperparameter values for the prior of $\tau$ such as HT($4, 1$), others suggest using prior information or the data to help inform this choice.  For example, \cite{Wakefield:2007, Wakefield:2009} suggests choosing hyperparameters based on an assumed range of the residual odds ratio.  In this approach the prior degrees of freedom is selected, and the scale parameter is then obtained under the assumed range with an IG prior for $\tau^2$. 

\cite{gelman2017prior} emphasized that a prior can only be understood and evaluated as it relates to a particular measurement process, i.e., a likelihood.  They discourage reliance on default principles such as maximum entropy or weak informativeness, supporting the abandonment of one-size-fits-all priors.  The authors argue that it is a false paradox that prior knowledge must exist without reference to the likelihood, but that the form of the prior itself must make sense with the likelihood. This apparent paradox is resolved by considering the choice of prior in a broader context of the entire analysis. There, the prior must be valid in terms of {\it a priori} data generation (i.e., via the prior predictive distribution) and in terms of data prediction (i.e., via the posterior predictive distribution). 

\cite{Simpson:2017} developed Penalized Complexity priors that follow a number of principles, including that the prior must not be noninformative.  Using the concept of a base model, which here would correspond to the model without random effects, they argue that the prior must place a sufficient mass at the base model, but also sufficient mass in the tails so that a more complex model is not over-penalized.  These principles would favor a HT prior over an IG prior, because the latter has insufficient mass at $\tau^2=0$.  However, \cite{Simpson:2017} noted that priors with a Gaussian tail can perform badly, suggesting that a HT prior with very large $\nu$ could have poor performance.  Their recommendation is for a prior with exponential tails.

There is a difference of opinion in the literature as to whether a more simple or more complex model should be favored {\it a priori}.  \cite{Barr:2013} argue for retaining all random effects that can be justified by the data in the context of confirmatory analyses when testing hypotheses about fixed effects.  In contrast, \cite{Simpson:2017}, \cite{Gustafson:2006} and others suggest selecting a prior that favors the simpler model.  \cite{Bates:2018} show the problems that can occur if a model is fit that is too complex for the data, including convergence to the boundary and singular covariance matrices.  \cite{Bates:2018} illustrate these issues in models that have multiple random effects that can potentially be correlated with each other.  In this paper we consider models with either a single random effect, or with two random effects that are independent due to the data structure.  Here we focus on making inference about $\tau$, but not testing whether $\tau=0$. 

  
  The focus of this work is on evaluating the performance of prior family and hyperparameter choice. We consider IG($a, b$) priors for $\tau^2$ or HT($\nu, s$) priors for $\tau$  over a range of hyperparameter values for three hierarchical models using simulated data.   Our evaluation is based on coverage of the true value of $\tau$ in simulated data, as well as average bias and length of the 95\% posterior interval.  \cite{BrowneDraper:2006} used similar criteria to evaluate IG($\epsilon, \epsilon$) and Unif($0, 1/\epsilon$) priors for $\tau^2$ with very small $\epsilon$ for two hierarchical models, arguing for the importance of priors that lead to well-calibrated inference.  In our evaluation we consider several sets of hyperparameters recommended in the recent Bayesian literature for both the IG and HT priors.  We also evaluate the performance of IG and HT priors using values that are determined in the context of the scale of the data.  
  
  %
  
\section{Methods}

\subsection{Models considered}

We consider the following three models:

\underline{Model 1}: Our first model is the eight-schools model, given in (\ref{eq:8schools1}) and (\ref{eq:8schools2}).  We assume that $\sigma^2_j$, $j=1, \cdots, J$ are known, and use the values of $\sigma^2_j$ from the original eight schools data.  Since estimation of the overall mean is not of interest here, in our simulated data each $\bar{y}$ has mean 0 and we do not estimate $\mu$.  Simulated data using $\tau=2.61$ approximately reproduces the variability of the $\bar{y}$'s in the original eight-schools data.  In our simulations we use three values of $\tau$: $\tau=0.4, 2$, and $10$. In all three models, the model specification is complete upon choosing the prior (IG or HT) for $\tau^2$ or $\tau$, and associated hyperparameters.  

\underline{Model 2}: Our second model is a longitudinal model in which we allow  subject-specific intercepts but assume a common slope across all subjects.  We chose not to allow subject-specific slopes so that we retain our focus on evaluating priors for scalar random effects, rather than considering an inverse-Wishart prior for covariance matrices.   Model 2 is
\begin{eqnarray}
  \label{eq:longit1}
  y_{i,j} & =&  \beta_0 + \beta_1 x_{i,j} + \alpha_i + \epsilon_{i,j}, \ i=1, \cdots, n, \ j=1, \cdots, J \\
  \alpha_i & \sim & N(0, \tau^2) 
\end{eqnarray}
where $y_{i,j}$ is the outcome at the $j$th age of the $i$th subject, and $x_{i,j}$ denotes the corresponding age.  In this model we assume that all subjects are measured at the same $J$ ages.  We also center $x_{i,j}$ so that $\beta_0 = 0$ on average.  As in the previous model, interest focuses on the prior for $\tau^2$.  To illustrate this model, we use $\beta_1=0.2$, $\sigma=1$,  $J=4$, $n=10$ or $30$, and again use three values of $\tau$: $\tau=0.1, 1$, or $10$.   In Model 2 we also use the following priors for the other model parameters
\begin{eqnarray*}
  \epsilon_{i,j}  \sim  N(0, \sigma^2), \ (\beta_0, \beta_1)^T \sim N(0,   10^2 \sigma^2  I), \  
  \sigma^2 \sim \mbox{IG}(0.05, 0.01)
\end{eqnarray*}

\underline{Model 3}:  Our final model is a multiple outcomes model with two scalar random effects.  In this model we assume that each of $n$ subjects has measurements of $J$ outcomes, and that the outcomes can be influenced by a single covariate.  We allow the slope of the covariate to differ across the $J$ outcomes, and treat the slope deviations as random effects with standard deviation $\tau_b$.   The model also includes a random subject effect, $r_i$ with scalar standard deviation $\tau_r$, which induces an exchangeable correlation between multiple outcomes measured on the same subjects. The model is:
\begin{eqnarray}
  \label{eq:multout1}
  y_{i,j} & = & \beta_0 + (\beta_1 + b_{j}) x_i + r_i + \epsilon_{i,j}, \ i=1, \cdots, n, \ j=1, \cdots, J \\
  \label{eq:multout2}
  b_j & \sim & N(0, \tau^2_b), \ 
  r_i  \sim N(0, \tau^2_r) \\
  \nonumber
    \epsilon_{i,j} & \sim & N(0, \sigma^2), \ (\beta_0, \beta_1)^T \sim N(0,   10^2 \sigma^2  I),  \
    \sigma^2 \sim IG(0.05, 0.01) 
\end{eqnarray}
  In simulations for Model 3 we use $\beta_0=0$, $\beta_1=0.2$, $\sigma^2=1$, and center and scale the vector of $x_i$ so that $y_{i,j}$ has mean $0$ on average.  All simulations use $n=700$ and $J=7$. We simulate data using $\tau_r^2=0.7$ and two values of $\tau_b$: $\tau_b = 0.04$ or $0.16$.  The values of $\tau_r$ and $\tau_b$ were chosen to be representative of multiple outcomes models that are based on models in (\ref{eq:multout1}), see \cite{Thurston:2009}.

\subsection{Choice of hyperparameters}

 {\bf Inverse gamma prior for $\tau^2$:}  We use the  IG($a,b$) parameterization with kernel $(\tau^2)^{-(a+1)} \mbox{exp}(-b/\tau^2)$, as in \cite{GelmanBDA:2013}.  The IG($a,b$) distribution is equivalent to a scaled inverse $\chi^2$ prior with scale $s$ and $\nu$ degrees of freedom and we can write the IG($a,b$) prior as  IG($\nu/2, s^2 \nu/2$). Taking the 8-schools model as an example for which $\tau^2$ is the variance of $\theta$, the posterior for $\tau^2$ is
\begin{equation}
  \label{eq:post}
  \tau^2 \mid \theta \sim IG\left(\frac{\nu + J}{2}, \frac{\nu s^2 +  \theta^T \theta}{2} \right) = IG \left(a + \frac{J}{2}, b + \frac{\theta^T \theta}{2} \right).
\end{equation}

Since $\nu = 2a$ adds to the total number of random effects ($J$) to give the posterior sample size, $\nu$ is the prior degrees of freedom, which is also the prior sample size for the number of random effects. Also $\nu s^2=2 b$ adds to $ \theta^T \theta$ to  give the posterior sum of squares, so $s^2$ is the prior variance and $\nu s^2$ is the prior sum of squares.  Under a goal of choosing a prior that does not overwhelm the information from the data, it would be reasonable to choose values of the hyperparameters that are somewhat small relative to the data.  We consider several choices of $\nu$ (or equivalently $a$) to evaluate the effect of changing only the prior degrees of freedom.  Our smallest value is $\nu = 1$ (i.e. $a=0.5$), so that the information in the prior is equivalent to a prior sample size of $1$ random effect.   Due to known convergence issues with very small degrees of freedom, our intermediate choice is $\nu=4$, which corresponds to  $a=2$.  Our final choice is $\nu=10$ (e.g. $a=5$).  

In terms of choosing reasonable values for  $b$, we see no reason to set $b=a$.  Instead, it is important to avoid picking $s^2$ (or equivalently, $b$) that is too large relative to the unknown value of $\tau^2$ in the data, a point noted by others \citep{Crainiceanu:2008}.  Since the posterior mean of $\tau^2$ is $(\nu s^2 + \theta^T \theta) / (\nu + J - 1)$, $\nu$ and $s^2$ can be chosen independently, while changing $a$ would result in changing $b$ to $b=a s^2$. If $\tau^2$ was known, we could consider choosing $s^2$ or $b$ so that $s^2 = \tau^2$ and $b = a  \tau^2$.

{\bf Half-t prior for $\tau$:} As shown by \cite{Wand:2011} and \cite{Huang:2013}, if $\tau^2 \mid a \sim IG(\nu/2, \nu/a)$ and $a \sim IG(1/2, 1/s^2)$, then $\tau \sim \mbox{HT}(\nu,s)$.  Therefore the IG prior for $\tau^2$ directly relates  to a HT prior for $\tau$ with $\nu = 2a$ and $s = \tau$.  Like the IG prior, $\nu$ in the HT prior is the {\it a priori} degrees of freedom, or prior number of random effects, and $s^2$ is the prior estimate for $\tau^2$.  Given the relationship between the IG and HT distributions, any values of $a$ and $b$ that we choose for an IG prior for $\tau^2$ can be directly translated into values of $\nu$ and $s$ for a HT prior for $\tau$.   We argue that for a given $\nu$ and $s$, the $HT(\nu, s)$ prior for $\tau$ in some sense has equivalent information regarding the prior degrees of freedom and prior scale parameter as the $IG(a=\nu/2, b=\nu s^2/2)$ prior for $\tau^2$.    When evaluating possible reasonable hyperparameter values, we consider IG and HT priors in pairs as defined here.  In what follows when referring to prior hyperparameters we use $\nu$ to denote the prior degrees of freedom (which is $2a$ for the IG prior), and use the term "prior scale hyperparameter" to refer to $s$ for the both the HT prior and IG prior for which $s=\sqrt{2b/\nu}$.  

{\bf Prior hyperparameters used here:} Several sets of our prior hyperparameters for $s$ and $b$ are based on the true value of $\tau$.  Since $\tau$ is unknown in real situations, we also choose $s$ and $b$ to be larger or smaller than the true $\tau$, by a factor of $c=1.5$.  We use the IG($a,b$) notation convention to specify IG prior hyperparameters.  We evaluate each of the three models under the following 14 sets of priors and hyperparameter values:
(1)  IG($1,1$); 
(2)  IG($0.001, 0.001$); 
(3)  half-Cauchy($1.2 \ \tau$) = HT($1, 1.2 \tau$);
(4)  HT($4,1$); 
(5)  IG($0.5, \tau^2 /2$); 
(6)  HT($1, \tau$); 
(7) IG($2, 2 \tau^2$); 
(8)  HT($4, \tau$);  
(9)   IG($5, 5 \tau^2$);  
(10)  HT($10, \tau$); 
(11) IG($2,  2 (c  \tau)^2$ ); 
(12)  HT($4,  c \tau$); 
(13) IG($2,  2 (\tau /c)^2 ) $); 
(14)  HT($4, \tau/c$).

  Priors 1--4 are suggested by the literature.  Priors 5--10 use the true value of $\tau$ in determining $s$ and $b$, and compare results due to varying $\nu$ from the small (prior 5--6), intermediate (priors 7--8) and larger (priors 9--10) values that we considered.  Finally, priors 11--14 evaluate the impact of choosing $s$ and $b$ to be larger (priors 11--12) or smaller (priors 13--14) than the true $\tau$ under the intermediate $\nu=4$.   By choosing $c=1.5$ as the under- and over-estimation factor for  $\tau$, $c^2=2.25$ is the under- or over-estimation factor for $\tau^2$. 

For each model and each true value of $\tau$, we are primarily interested in comparing the IG and HT prior ``pairs'' as defined earlier, in particular; (i)  the effect of increasing the prior degrees of freedom ($\nu$ or $a$) when the true $\tau$ is used to determine $s$ or $b$; and (ii) the effect of taking the prior scale hyperparameter (a function of $s$ or $b$) to be larger or smaller than $\tau$, under the intermediate value of $\nu=4$, e.g. $a=2$.  We often refer to priors by their numbers, P-1, $\ldots$, P-14. Our focus in (i) compares priors  P-5, P-7 and P-9, with P-6, P-8, and P-10, and focus (ii) compares P-7, P-11, and P-13, with  P-8, P-12, and P-14.

\subsection{Simulation details}

We simulated 100 datasets for each model under each value of $\tau$ (or $\tau_b$ and $\tau_r$ for Model 3).  We used Rstan~\cite{Rstan:pkg} in R~\cite{R:citation} for MCMC sampling using Stan~\cite{Stan:pkg} to fit the models using each of the 14 priors for all 100 datasets for each true parameter value(s).  Our MCMC samplers used 4 chains each with a warmup of $250$ iterations followed by $2500$ draws per chain, resulting in $10,000$ posterior draws per dataset.  Although in some situations a larger warmup could be justifiable, 200 iterations is generally considered a minimum adequate warmup period (see \url{https://statmodeling.stat.columbia.edu/2017/12/15/burn-vs-warm-iterative-simulation-algorithms/} (accessed 9/30/20).  Preliminary simulations for Model 2 (not shown) also showed no benefit of using a $2500$ iteration warmup instead of a $250$ iteration warmup per chain.  

The value of adapt\_delta used in running Stan models can be set by the user and impacts the step size of the MCMC sampler, with larger values corresponding to smaller steps.  Larger values can result in slower exploration of the parameter space, whereas smaller values closer to 0.8 may cause divergent transitions whereby the sampler diverges from the intended path \citep{Betancourt:2016}.  We initially considered the default value of adapt\_delta=0.8, but this resulted in a large number of divergent transitions for many priors over all three models.  Ultimately we chose adapt\_delta=0.99 for all models and all priors, to reduce to reduce the number of divergent transitions and make the results as comparable as possible across all conditions.  

For each simulation we report several convergence diagnostics, specifically the minimum effective sample size (ESS), median  ESS,  mean $\hat{R}$, and maximum $\hat{R}$. $\hat{R}$ is the "potential scale reduction factor" 
\citep{GelmanRubin:1992} that compares between chain and within chain variability, where smaller values close to $1$ are desirable.  We also report the mean number of divergent transitions (DT), the percentage of datasets with zero DT, and the maximum number of DTs.

Our evaluation of the priors and hyperparameters is based on the following, where $\tau$ indicates the true value and $\bar{\tau}$ indicates the mean over all MCMC draws for a given dataset:  the bias ($\bar{\tau} - \tau$), the relative bias $(\bar{\tau} - \tau)/\tau$, the root mean squared error (RMSE) $\sqrt{\mbox{MSE}}$, the coverage, and the length of the 95\% posterior interval.  Except for the RMSE, all quantities were calculated for each dataset and the stated summary statistics were calculated by averaging the results across all datasets.  The RMSE was calculated by averaging the mean squared error (MSE) $(\bar{\tau} - \tau)^2$ across all datasets, and taking the square root of the final averaged MSE \url{https://cran.r-project.org/web/packages/simhelpers/vignettes/MCSE.html} (accessed 3/28/21). 

\section{Results}

For each  model and true $\tau$ we start by commenting on convergence diagnostics.  In then discussing the performance of the priors, we first discuss the priors suggested by the literature (priors 1--4), then the effects of increasing the prior degrees of freedom, $\nu$ (priors 5--10), and finally the effects of over or underestimating the scale hyperparameter (priors 5-6 and 11-14).

\subsection{Model 1: the 8-schools model}

{\bf Small $\tau=0.4$:}   With $\tau = 0.4$, DTs occurred under two priors.  28\% of the datasets had at least one DT under the IG($0.001, 0.001$) (P-2) prior and a single dataset had one DT under the IG($0.5, \tau^2/2$) (P-5).   For all priors the mean $\hat{R}$ was $1.000$ and the median ESS was $\ge 4500$ (Table~\ref{tab:model1-diag}).  ESS was larger for the HT priors by a factor of 1.4 to 2.4 compared to their IG counterparts.

With the exception of the IG($1,1$) (P-1), the remaining priors had a coverage of 1.00.   The four priors recommended in the literature (priors 1--4) did not perform well. In addition to coverage of 0\%, the IG($1,1$) (P-1) and especially the IG($0.001, 0.001$) (P-2) were very badly biased (Table~\ref{tab:model1-estimates}).  Although the two HT priors recommended in the literature (P-3 and P-4) had better performance than P-1 and P-2, both the HT($1, 1.2 \tau$) (P-3) and HT($4,1$) (P-4) badly overestimated $\tau$, with a mean bias of $0.66$ and $0.59$ respectively.

\begin{table}[!tbp]
\centering\begingroup\fontsize{10}{12}\selectfont
\begin{tabular}[t]{lrrrrrrrr}
\toprule
  & \makecell[c]{True\\Value} & \makecell[c]{Min\\ESS} & \makecell[c]{Med\\ESS} & \makecell[c]{Mean\\Rhat} & \makecell[c]{Max\\Rhat} & \makecell[c]{Mean\\Div} & \makecell[c]{\% 0\\DT} & \makecell[c]{Max\\DT}\\
\midrule
\cellcolor{gray!6}{$1.IG(1, 1)$} & \cellcolor{gray!6}{0.4} & \cellcolor{gray!6}{2069.971} & \cellcolor{gray!6}{4869.495} & \cellcolor{gray!6}{1.000} & \cellcolor{gray!6}{1.003} & \cellcolor{gray!6}{0.00} & \cellcolor{gray!6}{100} & \cellcolor{gray!6}{0}\\
\cellcolor{gray!6}{$2.IG(0.001, 0.001)$} & \cellcolor{gray!6}{0.4} & \cellcolor{gray!6}{1227.432} & \cellcolor{gray!6}{5486.913} & \cellcolor{gray!6}{1.000} & \cellcolor{gray!6}{1.006} & \cellcolor{gray!6}{0.76} & \cellcolor{gray!6}{72} & \cellcolor{gray!6}{9}\\
$3.HT(1, 1.2\tau)$ & 0.4 & 1398.662 & 7991.351 & 1.000 & 1.003 & 0.00 & 100 & 0\\
$4.HT(4,1)$ & 0.4 & 7124.598 & 11989.097 & 1.000 & 1.001 & 0.00 & 100 & 0\\
\cellcolor{gray!6}{$5.IG(0.5, \tau ^2 / 2)$} & \cellcolor{gray!6}{0.4} & \cellcolor{gray!6}{948.101} & \cellcolor{gray!6}{4526.399} & \cellcolor{gray!6}{1.000} & \cellcolor{gray!6}{1.003} & \cellcolor{gray!6}{0.01} & \cellcolor{gray!6}{99} & \cellcolor{gray!6}{1}\\
\cellcolor{gray!6}{$6.HT(1, \tau)$} & \cellcolor{gray!6}{0.4} & \cellcolor{gray!6}{1151.190} & \cellcolor{gray!6}{7384.026} & \cellcolor{gray!6}{1.000} & \cellcolor{gray!6}{1.002} & \cellcolor{gray!6}{0.00} & \cellcolor{gray!6}{100} & \cellcolor{gray!6}{0}\\
$7.IG(2, 2 \tau ^2)$ & 0.4 & 2322.156 & 5291.816 & 1.000 & 1.002 & 0.00 & 100 & 0\\
$8.HT(4, \tau)$ & 0.4 & 3691.767 & 12474.579 & 1.000 & 1.001 & 0.00 & 100 & 0\\
\cellcolor{gray!6}{$9.IG(5, 5 \tau ^2)$} & \cellcolor{gray!6}{0.4} & \cellcolor{gray!6}{5803.598} & \cellcolor{gray!6}{8782.165} & \cellcolor{gray!6}{1.000} & \cellcolor{gray!6}{1.001} & \cellcolor{gray!6}{0.00} & \cellcolor{gray!6}{100} & \cellcolor{gray!6}{0}\\
\cellcolor{gray!6}{$10.HT(10, \tau)$} & \cellcolor{gray!6}{0.4} & \cellcolor{gray!6}{8750.311} & \cellcolor{gray!6}{12332.017} & \cellcolor{gray!6}{1.000} & \cellcolor{gray!6}{1.001} & \cellcolor{gray!6}{0.00} & \cellcolor{gray!6}{100} & \cellcolor{gray!6}{0}\\
$11.IG(2, 2(c \tau)^2)$ & 0.4 & 3041.727 & 5451.906 & 1.000 & 1.002 & 0.00 & 100 & 0\\
$12.HT(4, c \tau)$ & 0.4 & 8567.794 & 12043.162 & 1.000 & 1.001 & 0.00 & 100 & 0\\
\cellcolor{gray!6}{$13.IG(2, 2(\tau / c)^2)$} & \cellcolor{gray!6}{0.4} & \cellcolor{gray!6}{3193.071} & \cellcolor{gray!6}{5539.411} & \cellcolor{gray!6}{1.000} & \cellcolor{gray!6}{1.002} & \cellcolor{gray!6}{0.00} & \cellcolor{gray!6}{100} & \cellcolor{gray!6}{0}\\
\cellcolor{gray!6}{$14.HT(4, \tau / c)$} & \cellcolor{gray!6}{0.4} & \cellcolor{gray!6}{9343.910} & \cellcolor{gray!6}{12417.786} & \cellcolor{gray!6}{1.000} & \cellcolor{gray!6}{1.000} & \cellcolor{gray!6}{0.00} & \cellcolor{gray!6}{100} & \cellcolor{gray!6}{0}\\
\hline
$1.IG(1, 1)$ & 2.0 & 2008.961 & 4907.648 & 1.000 & 1.002 & 0.00 & 100 & 0\\
$2.IG(0.001, 0.001)$ & 2.0 & 1064.415 & 5172.395 & 1.001 & 1.008 & 0.66 & 68 & 9\\
\cellcolor{gray!6}{$3.HT(1, 1.2\tau)$} & \cellcolor{gray!6}{2.0} & \cellcolor{gray!6}{2350.516} & \cellcolor{gray!6}{8218.841} & \cellcolor{gray!6}{1.000} & \cellcolor{gray!6}{1.002} & \cellcolor{gray!6}{0.00} & \cellcolor{gray!6}{100} & \cellcolor{gray!6}{0}\\
\cellcolor{gray!6}{$4.HT(4,1)$} & \cellcolor{gray!6}{2.0} & \cellcolor{gray!6}{7554.063} & \cellcolor{gray!6}{11990.533} & \cellcolor{gray!6}{1.000} & \cellcolor{gray!6}{1.001} & \cellcolor{gray!6}{0.00} & \cellcolor{gray!6}{100} & \cellcolor{gray!6}{0}\\
$5.IG(0.5, \tau ^2 / 2)$ & 2.0 & 2157.761 & 5415.030 & 1.000 & 1.003 & 0.00 & 100 & 0\\
$6.HT(1, \tau)$ & 2.0 & 2604.746 & 8212.233 & 1.000 & 1.002 & 0.00 & 100 & 0\\
\cellcolor{gray!6}{$7.IG(2, 2 \tau ^2)$} & \cellcolor{gray!6}{2.0} & \cellcolor{gray!6}{2825.648} & \cellcolor{gray!6}{6079.181} & \cellcolor{gray!6}{1.000} & \cellcolor{gray!6}{1.001} & \cellcolor{gray!6}{0.00} & \cellcolor{gray!6}{100} & \cellcolor{gray!6}{0}\\
\cellcolor{gray!6}{$8.HT(4, \tau)$} & \cellcolor{gray!6}{2.0} & \cellcolor{gray!6}{5699.868} & \cellcolor{gray!6}{10646.604} & \cellcolor{gray!6}{1.000} & \cellcolor{gray!6}{1.001} & \cellcolor{gray!6}{0.00} & \cellcolor{gray!6}{100} & \cellcolor{gray!6}{0}\\
$9.IG(5, 5 \tau ^2)$ & 2.0 & 5879.421 & 8598.958 & 1.000 & 1.001 & 0.00 & 100 & 0\\
$10.HT(10, \tau)$ & 2.0 & 7155.451 & 10193.989 & 1.000 & 1.001 & 0.00 & 100 & 0\\
\cellcolor{gray!6}{$11.IG(2, 2(c \tau)^2)$} & \cellcolor{gray!6}{2.0} & \cellcolor{gray!6}{4233.307} & \cellcolor{gray!6}{6546.419} & \cellcolor{gray!6}{1.000} & \cellcolor{gray!6}{1.001} & \cellcolor{gray!6}{0.00} & \cellcolor{gray!6}{100} & \cellcolor{gray!6}{0}\\
\cellcolor{gray!6}{$12.HT(4, c \tau)$} & \cellcolor{gray!6}{2.0} & \cellcolor{gray!6}{4379.600} & \cellcolor{gray!6}{9142.929} & \cellcolor{gray!6}{1.000} & \cellcolor{gray!6}{1.001} & \cellcolor{gray!6}{0.00} & \cellcolor{gray!6}{100} & \cellcolor{gray!6}{0}\\
$13.IG(2, 2(\tau / c)^2)$ & 2.0 & 2448.457 & 6040.736 & 1.000 & 1.001 & 0.00 & 100 & 0\\
$14.HT(4, \tau / c)$ & 2.0 & 7427.532 & 11640.386 & 1.000 & 1.001 & 0.00 & 100 & 0\\
\hline
\cellcolor{gray!6}{$1.IG(1, 1)$} & \cellcolor{gray!6}{10.0} & \cellcolor{gray!6}{922.450} & \cellcolor{gray!6}{3899.285} & \cellcolor{gray!6}{1.001} & \cellcolor{gray!6}{1.008} & \cellcolor{gray!6}{0.00} & \cellcolor{gray!6}{100} & \cellcolor{gray!6}{0}\\
\cellcolor{gray!6}{$2.IG(0.001, 0.001)$} & \cellcolor{gray!6}{10.0} & \cellcolor{gray!6}{865.621} & \cellcolor{gray!6}{2695.280} & \cellcolor{gray!6}{1.001} & \cellcolor{gray!6}{1.008} & \cellcolor{gray!6}{1.26} & \cellcolor{gray!6}{75} & \cellcolor{gray!6}{55}\\
$3.HT(1, 1.2\tau)$ & 10.0 & 2101.793 & 3903.479 & 1.001 & 1.003 & 0.00 & 100 & 0\\
$4.HT(4,1)$ & 10.0 & 1347.729 & 10854.274 & 1.000 & 1.004 & 0.00 & 100 & 0\\
\cellcolor{gray!6}{$5.IG(0.5, \tau ^2 / 2)$} & \cellcolor{gray!6}{10.0} & \cellcolor{gray!6}{2522.406} & \cellcolor{gray!6}{4401.440} & \cellcolor{gray!6}{1.001} & \cellcolor{gray!6}{1.002} & \cellcolor{gray!6}{0.00} & \cellcolor{gray!6}{100} & \cellcolor{gray!6}{0}\\
\cellcolor{gray!6}{$6.HT(1, \tau)$} & \cellcolor{gray!6}{10.0} & \cellcolor{gray!6}{1848.880} & \cellcolor{gray!6}{3906.667} & \cellcolor{gray!6}{1.001} & \cellcolor{gray!6}{1.003} & \cellcolor{gray!6}{0.00} & \cellcolor{gray!6}{100} & \cellcolor{gray!6}{0}\\
$7.IG(2, 2 \tau ^2)$ & 10.0 & 4082.312 & 5575.471 & 1.000 & 1.002 & 0.00 & 100 & 0\\
$8.HT(4, \tau)$ & 10.0 & 2540.422 & 4304.376 & 1.001 & 1.003 & 0.00 & 100 & 0\\
\cellcolor{gray!6}{$9.IG(5, 5 \tau ^2)$} & \cellcolor{gray!6}{10.0} & \cellcolor{gray!6}{4780.271} & \cellcolor{gray!6}{7307.706} & \cellcolor{gray!6}{1.000} & \cellcolor{gray!6}{1.001} & \cellcolor{gray!6}{0.00} & \cellcolor{gray!6}{100} & \cellcolor{gray!6}{0}\\
\cellcolor{gray!6}{$10.HT(10, \tau)$} & \cellcolor{gray!6}{10.0} & \cellcolor{gray!6}{2548.573} & \cellcolor{gray!6}{4351.740} & \cellcolor{gray!6}{1.000} & \cellcolor{gray!6}{1.003} & \cellcolor{gray!6}{0.00} & \cellcolor{gray!6}{100} & \cellcolor{gray!6}{0}\\
$11.IG(2, 2(c \tau)^2)$ & 10.0 & 3210.570 & 5096.866 & 1.000 & 1.002 & 0.00 & 100 & 0\\
$12.HT(4, c \tau)$ & 10.0 & 2442.312 & 3916.177 & 1.001 & 1.003 & 0.00 & 100 & 0\\
\cellcolor{gray!6}{$13.IG(2, 2(\tau / c)^2)$} & \cellcolor{gray!6}{10.0} & \cellcolor{gray!6}{3346.894} & \cellcolor{gray!6}{6123.470} & \cellcolor{gray!6}{1.000} & \cellcolor{gray!6}{1.002} & \cellcolor{gray!6}{0.00} & \cellcolor{gray!6}{100} & \cellcolor{gray!6}{0}\\
\cellcolor{gray!6}{$14.HT(4, \tau / c)$} & \cellcolor{gray!6}{10.0} & \cellcolor{gray!6}{2526.741} & \cellcolor{gray!6}{4879.733} & \cellcolor{gray!6}{1.000} & \cellcolor{gray!6}{1.002} & \cellcolor{gray!6}{0.00} & \cellcolor{gray!6}{100} & \cellcolor{gray!6}{0}\\
\bottomrule
\end{tabular}
\endgroup{}
 \caption{\label{tab:model1-diag}Diagnostics for Model 1, the 8-schools model.}
\end{table}

\begin{table}[!tbp]
\centering\begingroup\fontsize{10}{12}\selectfont

\begin{tabular}[t]{lrrrrrrrr}
\toprule
  & \makecell[c]{True\\Value} & Mean & Median & Bias & \makecell[c]{Rel\\Bias} & RMSE & Cov & \makecell[c]{Interval\\Length}\\
\midrule
\cellcolor{gray!6}{$1.IG(1, 1)$} & \cellcolor{gray!6}{0.4} & \cellcolor{gray!6}{1.619} & \cellcolor{gray!6}{1.193} & \cellcolor{gray!6}{1.219} & \cellcolor{gray!6}{3.047} & \cellcolor{gray!6}{1.242} & \cellcolor{gray!6}{0.00} & \cellcolor{gray!6}{4.925}\\
\cellcolor{gray!6}{$2.IG(0.001, 0.001)$} & \cellcolor{gray!6}{0.4} & \cellcolor{gray!6}{2.463} & \cellcolor{gray!6}{1.007} & \cellcolor{gray!6}{2.063} & \cellcolor{gray!6}{5.156} & \cellcolor{gray!6}{2.646} & \cellcolor{gray!6}{1.00} & \cellcolor{gray!6}{12.287}\\
$3.HT(1, 1.2\tau)$ & 0.4 & 1.062 & 0.466 & 0.662 & 1.656 & 0.804 & 1.00 & 6.081\\
$4.HT(4,1)$ & 0.4 & 0.988 & 0.737 & 0.588 & 1.470 & 0.591 & 1.00 & 3.405\\
\cellcolor{gray!6}{$5.IG(0.5, \tau ^2 / 2)$} & \cellcolor{gray!6}{0.4} & \cellcolor{gray!6}{1.185} & \cellcolor{gray!6}{0.579} & \cellcolor{gray!6}{0.785} & \cellcolor{gray!6}{1.963} & \cellcolor{gray!6}{0.907} & \cellcolor{gray!6}{1.00} & \cellcolor{gray!6}{6.021}\\
\cellcolor{gray!6}{$6.HT(1, \tau)$} & \cellcolor{gray!6}{0.4} & \cellcolor{gray!6}{0.934} & \cellcolor{gray!6}{0.389} & \cellcolor{gray!6}{0.534} & \cellcolor{gray!6}{1.334} & \cellcolor{gray!6}{0.680} & \cellcolor{gray!6}{1.00} & \cellcolor{gray!6}{5.573}\\
$7.IG(2, 2 \tau ^2)$ & 0.4 & 0.501 & 0.437 & 0.101 & 0.254 & 0.102 & 1.00 & 0.911\\
$8.HT(4, \tau)$ & 0.4 & 0.400 & 0.297 & 0.000 & 0.000 & 0.006 & 1.00 & 1.381\\
\cellcolor{gray!6}{$9.IG(5, 5 \tau ^2)$} & \cellcolor{gray!6}{0.4} & \cellcolor{gray!6}{0.433} & \cellcolor{gray!6}{0.414} & \cellcolor{gray!6}{0.033} & \cellcolor{gray!6}{0.083} & \cellcolor{gray!6}{0.033} & \cellcolor{gray!6}{1.00} & \cellcolor{gray!6}{0.422}\\
\cellcolor{gray!6}{$10.HT(10, \tau)$} & \cellcolor{gray!6}{0.4} & \cellcolor{gray!6}{0.346} & \cellcolor{gray!6}{0.280} & \cellcolor{gray!6}{-0.054} & \cellcolor{gray!6}{-0.135} & \cellcolor{gray!6}{0.054} & \cellcolor{gray!6}{1.00} & \cellcolor{gray!6}{1.040}\\
$11.IG(2, 2(c \tau)^2)$ & 0.4 & 0.750 & 0.654 & 0.350 & 0.876 & 0.350 & 1.00 & 1.357\\
$12.HT(4, c \tau)$ & 0.4 & 0.597 & 0.444 & 0.197 & 0.493 & 0.198 & 1.00 & 2.057\\
\cellcolor{gray!6}{$13.IG(2, 2(\tau / c)^2)$} & \cellcolor{gray!6}{0.4} & \cellcolor{gray!6}{0.334} & \cellcolor{gray!6}{0.291} & \cellcolor{gray!6}{-0.066} & \cellcolor{gray!6}{-0.166} & \cellcolor{gray!6}{0.066} & \cellcolor{gray!6}{1.00} & \cellcolor{gray!6}{0.604}\\
\cellcolor{gray!6}{$14.HT(4, \tau / c)$} & \cellcolor{gray!6}{0.4} & \cellcolor{gray!6}{0.267} & \cellcolor{gray!6}{0.198} & \cellcolor{gray!6}{-0.133} & \cellcolor{gray!6}{-0.333} & \cellcolor{gray!6}{0.133} & \cellcolor{gray!6}{1.00} & \cellcolor{gray!6}{0.922}\\
\hline
$1.IG(1, 1)$ & 2.0 & 1.647 & 1.200 & -0.353 & -0.176 & 0.438 & 1.00 & 5.131\\
$2.IG(0.001, 0.001)$ & 2.0 & 2.657 & 1.109 & 0.657 & 0.329 & 1.786 & 1.00 & 12.995\\
\cellcolor{gray!6}{$3.HT(1, 1.2\tau)$} & \cellcolor{gray!6}{2.0} & \cellcolor{gray!6}{3.083} & \cellcolor{gray!6}{2.127} & \cellcolor{gray!6}{1.083} & \cellcolor{gray!6}{0.542} & \cellcolor{gray!6}{1.576} & \cellcolor{gray!6}{1.00} & \cellcolor{gray!6}{11.311}\\
\cellcolor{gray!6}{$4.HT(4,1)$} & \cellcolor{gray!6}{2.0} & \cellcolor{gray!6}{0.997} & \cellcolor{gray!6}{0.741} & \cellcolor{gray!6}{-1.003} & \cellcolor{gray!6}{-0.502} & \cellcolor{gray!6}{1.005} & \cellcolor{gray!6}{1.00} & \cellcolor{gray!6}{3.453}\\
$5.IG(0.5, \tau ^2 / 2)$ & 2.0 & 3.647 & 2.695 & 1.647 & 0.823 & 1.972 & 1.00 & 10.787\\
$6.HT(1, \tau)$ & 2.0 & 2.786 & 1.832 & 0.786 & 0.393 & 1.306 & 1.00 & 10.716\\
\cellcolor{gray!6}{$7.IG(2, 2 \tau ^2)$} & \cellcolor{gray!6}{2.0} & \cellcolor{gray!6}{2.485} & \cellcolor{gray!6}{2.183} & \cellcolor{gray!6}{0.485} & \cellcolor{gray!6}{0.243} & \cellcolor{gray!6}{0.509} & \cellcolor{gray!6}{1.00} & \cellcolor{gray!6}{4.418}\\
\cellcolor{gray!6}{$8.HT(4, \tau)$} & \cellcolor{gray!6}{2.0} & \cellcolor{gray!6}{1.943} & \cellcolor{gray!6}{1.479} & \cellcolor{gray!6}{-0.057} & \cellcolor{gray!6}{-0.029} & \cellcolor{gray!6}{0.311} & \cellcolor{gray!6}{1.00} & \cellcolor{gray!6}{6.458}\\
$9.IG(5, 5 \tau ^2)$ & 2.0 & 2.166 & 2.069 & 0.166 & 0.083 & 0.168 & 1.00 & 2.102\\
$10.HT(10, \tau)$ & 2.0 & 1.722 & 1.403 & -0.278 & -0.139 & 0.317 & 1.00 & 5.116\\
\cellcolor{gray!6}{$11.IG(2, 2(c \tau)^2)$} & \cellcolor{gray!6}{2.0} & \cellcolor{gray!6}{3.643} & \cellcolor{gray!6}{3.250} & \cellcolor{gray!6}{1.643} & \cellcolor{gray!6}{0.821} & \cellcolor{gray!6}{1.671} & \cellcolor{gray!6}{1.00} & \cellcolor{gray!6}{6.021}\\
\cellcolor{gray!6}{$12.HT(4, c \tau)$} & \cellcolor{gray!6}{2.0} & \cellcolor{gray!6}{2.775} & \cellcolor{gray!6}{2.196} & \cellcolor{gray!6}{0.775} & \cellcolor{gray!6}{0.387} & \cellcolor{gray!6}{0.976} & \cellcolor{gray!6}{1.00} & \cellcolor{gray!6}{8.650}\\
$13.IG(2, 2(\tau / c)^2)$ & 2.0 & 1.664 & 1.455 & -0.336 & -0.168 & 0.341 & 1.00 & 3.000\\
$14.HT(4, \tau / c)$ & 2.0 & 1.322 & 0.988 & -0.678 & -0.339 & 0.691 & 1.00 & 4.549\\
\hline
\cellcolor{gray!6}{$1.IG(1, 1)$} & \cellcolor{gray!6}{10.0} & \cellcolor{gray!6}{2.360} & \cellcolor{gray!6}{1.662} & \cellcolor{gray!6}{-7.640} & \cellcolor{gray!6}{-0.764} & \cellcolor{gray!6}{7.895} & \cellcolor{gray!6}{0.20} & \cellcolor{gray!6}{7.987}\\
\cellcolor{gray!6}{$2.IG(0.001, 0.001)$} & \cellcolor{gray!6}{10.0} & \cellcolor{gray!6}{5.247} & \cellcolor{gray!6}{3.413} & \cellcolor{gray!6}{-4.753} & \cellcolor{gray!6}{-0.475} & \cellcolor{gray!6}{6.766} & \cellcolor{gray!6}{0.88} & \cellcolor{gray!6}{18.701}\\
$3.HT(1, 1.2\tau)$ & 10.0 & 8.642 & 7.771 & -1.358 & -0.136 & 4.112 & 1.00 & 20.705\\
$4.HT(4,1)$ & 10.0 & 1.137 & 0.780 & -8.863 & -0.886 & 8.870 & 0.05 & 4.466\\
\cellcolor{gray!6}{$5.IG(0.5, \tau ^2 / 2)$} & \cellcolor{gray!6}{10.0} & \cellcolor{gray!6}{10.758} & \cellcolor{gray!6}{9.814} & \cellcolor{gray!6}{0.758} & \cellcolor{gray!6}{0.076} & \cellcolor{gray!6}{2.904} & \cellcolor{gray!6}{1.00} & \cellcolor{gray!6}{18.069}\\
\cellcolor{gray!6}{$6.HT(1, \tau)$} & \cellcolor{gray!6}{10.0} & \cellcolor{gray!6}{8.220} & \cellcolor{gray!6}{7.334} & \cellcolor{gray!6}{-1.780} & \cellcolor{gray!6}{-0.178} & \cellcolor{gray!6}{4.206} & \cellcolor{gray!6}{1.00} & \cellcolor{gray!6}{20.110}\\
$7.IG(2, 2 \tau ^2)$ & 10.0 & 10.826 & 10.213 & 0.826 & 0.083 & 1.770 & 1.00 & 13.150\\
$8.HT(4, \tau)$ & 10.0 & 7.955 & 7.262 & -2.045 & -0.205 & 3.903 & 1.00 & 18.488\\
\cellcolor{gray!6}{$9.IG(5, 5 \tau ^2)$} & \cellcolor{gray!6}{10.0} & \cellcolor{gray!6}{10.502} & \cellcolor{gray!6}{10.159} & \cellcolor{gray!6}{0.502} & \cellcolor{gray!6}{0.050} & \cellcolor{gray!6}{0.919} & \cellcolor{gray!6}{1.00} & \cellcolor{gray!6}{8.962}\\
\cellcolor{gray!6}{$10.HT(10, \tau)$} & \cellcolor{gray!6}{10.0} & \cellcolor{gray!6}{7.795} & \cellcolor{gray!6}{7.190} & \cellcolor{gray!6}{-2.205} & \cellcolor{gray!6}{-0.220} & \cellcolor{gray!6}{3.782} & \cellcolor{gray!6}{1.00} & \cellcolor{gray!6}{17.658}\\
$11.IG(2, 2(c \tau)^2)$ & 10.0 & 14.276 & 13.595 & 4.276 & 0.428 & 4.538 & 0.95 & 15.411\\
$12.HT(4, c \tau)$ & 10.0 & 9.141 & 8.362 & -0.859 & -0.086 & 3.836 & 1.00 & 20.849\\
\cellcolor{gray!6}{$13.IG(2, 2(\tau / c)^2)$} & \cellcolor{gray!6}{10.0} & \cellcolor{gray!6}{8.140} & \cellcolor{gray!6}{7.546} & \cellcolor{gray!6}{-1.860} & \cellcolor{gray!6}{-0.186} & \cellcolor{gray!6}{2.436} & \cellcolor{gray!6}{1.00} & \cellcolor{gray!6}{11.367}\\
\cellcolor{gray!6}{$14.HT(4, \tau / c)$} & \cellcolor{gray!6}{10.0} & \cellcolor{gray!6}{6.568} & \cellcolor{gray!6}{5.900} & \cellcolor{gray!6}{-3.432} & \cellcolor{gray!6}{-0.343} & \cellcolor{gray!6}{4.488} & \cellcolor{gray!6}{1.00} & \cellcolor{gray!6}{15.975}\\
\bottomrule
\end{tabular}
\endgroup{}
\caption{Summary results for $\tau$ under Model 1, the 8-schools model.   \label{tab:model1-estimates}}
\end{table}

\begin{figure}[tbp]
\includegraphics[width=6in,height=6in]{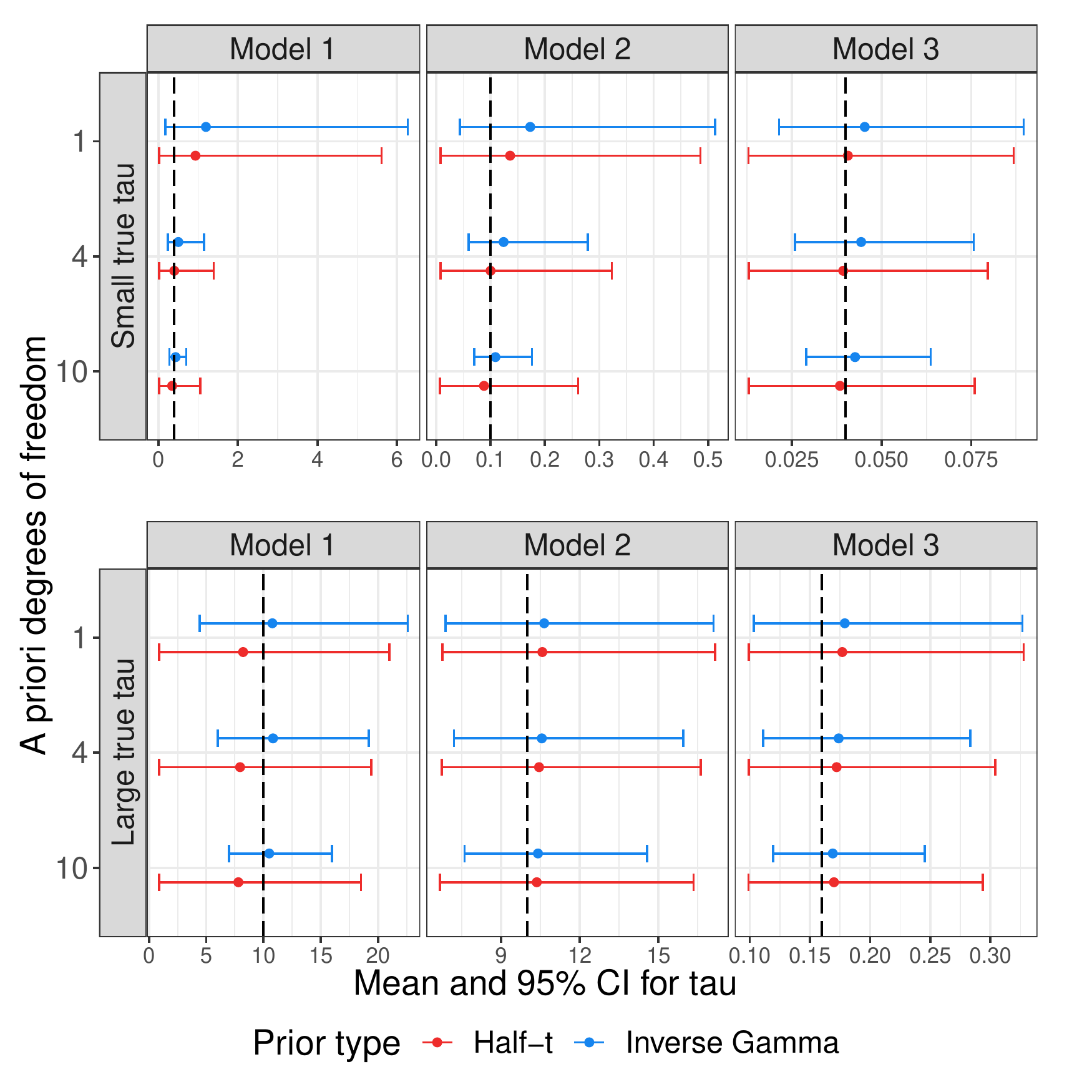}
\caption{Influence of {\it a priori} degrees of freedom, $\nu$, on posterior means and 95\% posterior intervals for $\tau$  under inverse gamma and half-t priors.  All priors shown use the true value of $\tau$ as the scale hyperparameter.  Results are organized by smallest or largest values of true $\tau$ (rows) and by model number (columns).  Vertical dashed lines indicate the true values of $\tau$.} 
\label{fig:figure1}
\end{figure}

Among priors using the true $\tau$ as the scale hyperparameter (priors 5-10), the two with the smallest $\nu=1$ had a very large bias of 0.785 for the IG (P-5) and 0.534 for the HT (P-6).   Bias generally decreased with increasing $\nu$, for both IG and HT priors.  For these six priors the HT had smaller RMSE than their IG counterparts, and a shorter interval length for small $\nu$ (P-6 versus P-5) but a longer interval length with both the medium (P-8 vs P-7) and large $\nu$ (P-10 vs P-9).  These relationships are shown in the top left panel of Figure~\ref{fig:figure1}.

\begin{figure}[tbp]
\includegraphics[width=6in,height=6in]{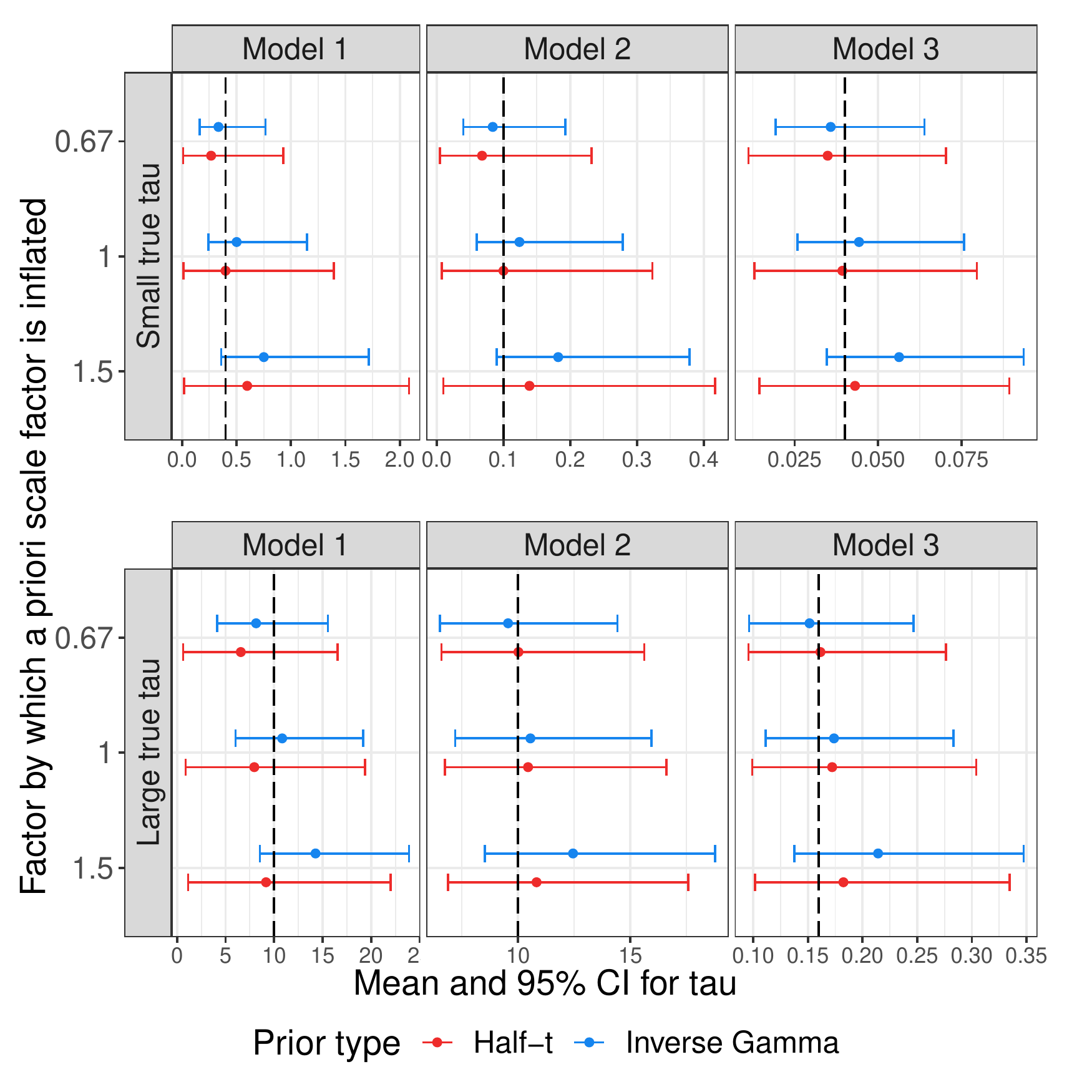}
\caption{Influence of choosing an {\it a priori} scale parameter that is smaller, equal, or larger to the true $\tau$ on posterior means and 95\% posterior intervals for $\tau$ under inverse gamma and half-t priors.  All priors shown use an {\it a priori} degrees of freedom of $\nu=4$.  Results are organized by smallest or largest true $\tau$ (rows) and by model number (columns).  Vertical dashed lines indicate the true values of $\tau$.}
\label{fig:figure2}
\end{figure}

The priors that used an inflated scale hyperparameter overestimated the posterior mean, where the bias for the IG prior (P-11) was substantially larger than its HT counterpart (P-12), while having a shorter interval length.  When the prior scale hyperparameter underestimates $\tau$, both priors had negative bias, but the IG (P-13) was less biased than its HT counterpart (P-14), as well as having a smaller RMSE and  smaller interval length (Figure~\ref{fig:figure2}, top left panel). 

{\bf Medium $\tau=2$:}  All convergence diagnostics were reasonable except the IG($0.001, 0.001$) (P-2) for which 32\% of the datasets had a small number of DTs.  Among priors 5--14 ESS was larger for the HT priors than their IG counterparts, by a factor of 1.19 to 1.93.  All priors for $\tau=2$ had a coverage of 1.00 in this simulation.  Among the four priors suggested by the literature,  the IG($1,1$) (P-1) and HT($4,1$) (P-4) both underestimated $\tau$, whereas the IG($0.001, 0.001$) (P-2) and HT($1, 1.2 \tau$) (P-3) overestimated $\tau$, where the absolute bias was larger for the HT than the IG priors.

Among priors using $\tau$ as the scale hyperparameter (priors 5--10), the IG priors had larger absolute bias and larger RMSE than their HT counterparts, except for the largest $\nu$ (P-9, P-10).    Priors using the smallest $\nu$ (P-5, P-6), were particularly biased, especially the IG($0.5, \tau^2/2$).  Although the IG and HT priors with small $\nu=1$ had a similar interval length, as $\nu$ increased the interval length for the IG prior became substantially shorter relative to its HT counterpart.  Both priors that used an inflated scale parameter had a positive bias that was larger for the IG($2, 2 (c \tau)^2$), P-11, than its HT counterpart, P-12.  In contrast, both priors that used a small scale hyperparameter had a negative bias that was larger in absolute value for the HT($4, \tau/c$), P-14, than its IG counterpart, P-13.

{\bf Large $\tau=10$}: Only the IG($1,1$) (P-1) had DTs, affecting 25\% of the datasets.  All priors had a mean $\hat{R}$ of $1.001$ or less and ESS $>2500$, which was generally smaller than for small or medium true values ($\tau=0.4$ or $\tau=2$). In contrast to small or medium $\tau$, with $\tau=10$ ESS for the IG priors were larger by a factor of 1.13 to 1.68 than their HT counterparts.

Three of the four priors suggested by the literature performed very poorly.  The IG($1,1$) prior (P-1) had 20\% coverage and a bias of $-7.64$, the HT($4,1$) prior (P-4) had 5\% coverage and a bias of $-8.86$, and the IG($0.001, 0.001$) (P-2) prior had 88\% coverage and a bias of $-4.75$.

The six priors using $\tau$ as the scale hyperparameter (priors 5--10) all had a coverage of 1.00, but the three IG priors had better performance in all aspects than their HT counterparts.  The three IG priors overestimated $\tau$, but by a smaller degree than the three HT priors underestimated it.   For example, the bias for the IG($2, 2 \tau^2$) (P-7)  and its HT($4, \tau$) counterpart (P-8) was 0.83 and -2.045 respectively.  For these six priors the interval length was larger for the HT than the IG priors with a widening gap in interval length as $\nu$ increased, reaching a doubling in interval length with $\nu=10$ (P-10 vs P-9) (Figure~\ref{fig:figure1}, lower left).  When the prior scale hyperparameter was overestimated, the IG prior (P-11) substantially overestimated $\tau$ on average while still maintaining 95\% coverage, whereas the HT prior (P-12) had relatively small bias and 100\% coverage.  When the prior scale hyperparameter was underestimated, both priors underestimated $\tau$, but the HT prior (P-14) had larger absolute bias, larger RMSE, and longer interval length than its IG counterpart (P-13) (Figure~\ref{fig:figure2}, lower left).

\subsection{Model 2: the longitudinal model}

Although $\tau$ is the parameter of primary interest, we also calculated performance diagnostics for the overall slope, $\beta_1$.  For each of the six combinations of $\tau=0.1, 1, \mbox{~or~}10$ and data with $n=10 \mbox{~or~}30$ subjects,  all 14 priors gave nearly identical estimates of $\beta_1$, and this parameter was estimated with very little bias and good coverage. For each data set we also estimated $\tau$ from a random effects model using \texttt{lme4} with the formula \texttt{Y $\sim$ X + (1 $\vert$ Subject)}. When $\tau = 0.1$, \texttt{lme4} had difficulty estimating $\tau$, where $51\% ~(37\%)$ of the models estimated $\hat{\tau} = 0$ when $n = 10 ~(n=30)$, and $\tau$ was overestimated on average (Supplemental Table~\ref{tab:lme4}).   For $\tau=1$ and $\tau=10$, $\hat{\tau}$ was never zero, and tau was slightly underestimated.  We now focus on estimating $\tau$ in our Bayesian models.

{\bf Small $\tau=0.1$}: Convergence diagnostics with $\tau=0.1$ indicated that Stan had difficulty sampling from the posterior especially for the HT priors.  For the smaller $n=10$ datasets all HT priors had DTs for 79\% to 92\% of the datasets.  The only IG prior with any DTs was $IG(0.001, 0.001)$ (P-2).  (Table \ref{tab:model2-diag-n10}).   The IG priors also had larger ESS and smaller $\hat{R}$ than the HT priors. Among priors that use the true $\tau$ as the scale hyperparameter, the median ESS was larger for the IG priors by a factor of $2.1$ (P-5 vs P-6; $\nu=1$), to $8.2$ (P-9 vs P-10; $\nu=10$).
In simulations using data with $n=30$ observations, all HT priors had some DTs, but a smaller proportion of datasets had DTs than for $n=10$.  However, with the exception of the $IG(1,1)$ prior, ESS and $\hat{R}$ were considerably worse with $n=30$  (Supplemental Table~\ref{tab:model2-diag-n30}) than $n=10$.

With $n=10$, the four priors we evaluated that were suggested in the literature (P-1 to P-4) generally did not perform well (Table~\ref{tab:model2-estimates-n10}).  The $IG(1,1)$ prior  badly overestimated $\tau$, and had a coverage of $0$.  The $HT(4,1)$ prior, while having good coverage, was badly biased upwards with the largest bias of the HT priors  and largest interval length.  The situation was similar with $n=30$, although biases were smaller (Supplemental Table~\ref{tab:model2-estimates-n30}.  With both $n=10$ and $n=30$, coverage for all but the IG(1,1) prior was at least 98\%.

\begin{table}[tbp]
\centering
\fontsize{10}{12}\selectfont
\begin{tabular}[t]{lrrrrrrrr}
\toprule
Prior & \makecell[c]{True\\Value} & \makecell[c]{Min\\ESS} & \makecell[c]{Med\\ESS} & \makecell[c]{Mean\\Rhat} & \makecell[c]{Max\\Rhat} & \makecell[c]{Mean\\Div} & \makecell[c]{\% 0\\DT} & \makecell[c]{Max\\DT}\\
\midrule
\cellcolor{gray!6}{$1.IG(1,1)$} & \cellcolor{gray!6}{0.1} & \cellcolor{gray!6}{4396} & \cellcolor{gray!6}{5516} & \cellcolor{gray!6}{1.000} & \cellcolor{gray!6}{1.002} & \cellcolor{gray!6}{0.00} & \cellcolor{gray!6}{100} & \cellcolor{gray!6}{0}\\
\cellcolor{gray!6}{$2.IG(.001,.001)$} & \cellcolor{gray!6}{0.1} & \cellcolor{gray!6}{361} & \cellcolor{gray!6}{569} & \cellcolor{gray!6}{1.007} & \cellcolor{gray!6}{1.017} & \cellcolor{gray!6}{0.25} & \cellcolor{gray!6}{86} & \cellcolor{gray!6}{4}\\
$3.HT(1,1.2\tau)$ & 0.1 & 11 & 397 & 1.013 & 1.148 & 15.86 & 8 & 391\\
$4.HT(4,1)$ & 0.1 & 6 & 564 & 1.010 & 1.280 & 14.03 & 21 & 1022\\
\cellcolor{gray!6}{$5.IG(1/2,\tau^2/2)$} & \cellcolor{gray!6}{0.1} & \cellcolor{gray!6}{427} & \cellcolor{gray!6}{811} & \cellcolor{gray!6}{1.005} & \cellcolor{gray!6}{1.015} & \cellcolor{gray!6}{0.00} & \cellcolor{gray!6}{100} & \cellcolor{gray!6}{0}\\
\cellcolor{gray!6}{$6.HT(1,\tau)$} & \cellcolor{gray!6}{0.1} & \cellcolor{gray!6}{10} & \cellcolor{gray!6}{379} & \cellcolor{gray!6}{1.013} & \cellcolor{gray!6}{1.160} & \cellcolor{gray!6}{37.01} & \cellcolor{gray!6}{7} & \cellcolor{gray!6}{1350}\\
$7.IG(2,2\tau^2)$ & 0.1 & 844 & 1607 & 1.002 & 1.007 & 0.00 & 100 & 0\\
$8.HT(4,\tau)$ & 0.1 & 36 & 426 & 1.012 & 1.150 & 10.71 & 13 & 169\\
\cellcolor{gray!6}{$9.IG(5,5\tau^2)$} & \cellcolor{gray!6}{0.1} & \cellcolor{gray!6}{2370} & \cellcolor{gray!6}{3433} & \cellcolor{gray!6}{1.001} & \cellcolor{gray!6}{1.003} & \cellcolor{gray!6}{0.00} & \cellcolor{gray!6}{100} & \cellcolor{gray!6}{0}\\
\cellcolor{gray!6}{$10.HT(10,\tau)$} & \cellcolor{gray!6}{0.1} & \cellcolor{gray!6}{7} & \cellcolor{gray!6}{419} & \cellcolor{gray!6}{1.016} & \cellcolor{gray!6}{1.224} & \cellcolor{gray!6}{31.06} & \cellcolor{gray!6}{12} & \cellcolor{gray!6}{808}\\
$11.IG(2,2(c\tau)^2)$ & 0.1 & 1053 & 1997 & 1.002 & 1.004 & 0.00 & 100 & 0\\
$12.HT(4,c\tau)$ & 0.1 & 8 & 412 & 1.013 & 1.214 & 34.57 & 13 & 1256\\
\cellcolor{gray!6}{$13.IG(2,2(\tau/c)^2)$} & \cellcolor{gray!6}{0.1} & \cellcolor{gray!6}{694} & \cellcolor{gray!6}{1402} & \cellcolor{gray!6}{1.003} & \cellcolor{gray!6}{1.008} & \cellcolor{gray!6}{0.00} & \cellcolor{gray!6}{100} & \cellcolor{gray!6}{0}\\
\cellcolor{gray!6}{$14.HT(4,\tau/c)$} & \cellcolor{gray!6}{0.1} & \cellcolor{gray!6}{10} & \cellcolor{gray!6}{423} & \cellcolor{gray!6}{1.014} & \cellcolor{gray!6}{1.172} & \cellcolor{gray!6}{41.19} & \cellcolor{gray!6}{8} & \cellcolor{gray!6}{1304}\\
\midrule
\cellcolor{gray!6}{$1.IG(1,1)$} & \cellcolor{gray!6}{1.0} & \cellcolor{gray!6}{3718} & \cellcolor{gray!6}{4953} & \cellcolor{gray!6}{1.000} & \cellcolor{gray!6}{1.002} & \cellcolor{gray!6}{0.00} & \cellcolor{gray!6}{100} & \cellcolor{gray!6}{0}\\
\cellcolor{gray!6}{$2.IG(.001,.001)$} & \cellcolor{gray!6}{1.0} & \cellcolor{gray!6}{31} & \cellcolor{gray!6}{3012} & \cellcolor{gray!6}{1.004} & \cellcolor{gray!6}{1.068} & \cellcolor{gray!6}{2.37} & \cellcolor{gray!6}{84} & \cellcolor{gray!6}{118}\\
$3.HT(1,1.2\tau)$ & 1.0 & 405 & 3611 & 1.001 & 1.011 & 0.15 & 93 & 4\\
$4.HT(4,1)$ & 1.0 & 602 & 3924 & 1.001 & 1.010 & 0.19 & 91 & 5\\
\cellcolor{gray!6}{$5.IG(1/2,\tau^2/2)$} & \cellcolor{gray!6}{1.0} & \cellcolor{gray!6}{1640} & \cellcolor{gray!6}{4637} & \cellcolor{gray!6}{1.001} & \cellcolor{gray!6}{1.004} & \cellcolor{gray!6}{0.00} & \cellcolor{gray!6}{100} & \cellcolor{gray!6}{0}\\
\cellcolor{gray!6}{$6.HT(1,\tau)$} & \cellcolor{gray!6}{1.0} & \cellcolor{gray!6}{576} & \cellcolor{gray!6}{3799} & \cellcolor{gray!6}{1.001} & \cellcolor{gray!6}{1.008} & \cellcolor{gray!6}{0.14} & \cellcolor{gray!6}{93} & \cellcolor{gray!6}{6}\\
$7.IG(2,2\tau^2)$ & 1.0 & 3315 & 5592 & 1.000 & 1.002 & 0.00 & 100 & 0\\
$8.HT(4,\tau)$ & 1.0 & 602 & 3924 & 1.001 & 1.010 & 0.19 & 91 & 5\\
\cellcolor{gray!6}{$9.IG(5,5\tau^2)$} & \cellcolor{gray!6}{1.0} & \cellcolor{gray!6}{4772} & \cellcolor{gray!6}{6603} & \cellcolor{gray!6}{1.000} & \cellcolor{gray!6}{1.002} & \cellcolor{gray!6}{0.00} & \cellcolor{gray!6}{100} & \cellcolor{gray!6}{0}\\
\cellcolor{gray!6}{$10.HT(10,\tau)$} & \cellcolor{gray!6}{1.0} & \cellcolor{gray!6}{650} & \cellcolor{gray!6}{4166} & \cellcolor{gray!6}{1.001} & \cellcolor{gray!6}{1.006} & \cellcolor{gray!6}{0.14} & \cellcolor{gray!6}{94} & \cellcolor{gray!6}{5}\\
$11.IG(2,2(c\tau)^2)$ & 1.0 & 3767 & 5651 & 1.000 & 1.002 & 0.00 & 100 & 0\\
$12.HT(4,c\tau)$ & 1.0 & 644 & 3939 & 1.001 & 1.010 & 0.12 & 94 & 4\\
\cellcolor{gray!6}{$13.IG(2,2(\tau/c)^2)$} & \cellcolor{gray!6}{1.0} & \cellcolor{gray!6}{3254} & \cellcolor{gray!6}{5129} & \cellcolor{gray!6}{1.000} & \cellcolor{gray!6}{1.003} & \cellcolor{gray!6}{0.00} & \cellcolor{gray!6}{100} & \cellcolor{gray!6}{0}\\
\cellcolor{gray!6}{$14.HT(4,\tau/c)$} & \cellcolor{gray!6}{1.0} & \cellcolor{gray!6}{302} & \cellcolor{gray!6}{3559} & \cellcolor{gray!6}{1.002} & \cellcolor{gray!6}{1.011} & \cellcolor{gray!6}{0.58} & \cellcolor{gray!6}{88} & \cellcolor{gray!6}{25}\\
\midrule
\cellcolor{gray!6}{$1.IG(1,1)$} & \cellcolor{gray!6}{10} & \cellcolor{gray!6}{1842} & \cellcolor{gray!6}{2640} & \cellcolor{gray!6}{1.001} & \cellcolor{gray!6}{1.003} & \cellcolor{gray!6}{0} & \cellcolor{gray!6}{100} & \cellcolor{gray!6}{0}\\
\cellcolor{gray!6}{$2.IG(.001,.001)$} & \cellcolor{gray!6}{10} & \cellcolor{gray!6}{600} & \cellcolor{gray!6}{2428} & \cellcolor{gray!6}{1.001} & \cellcolor{gray!6}{1.007} & \cellcolor{gray!6}{0} & \cellcolor{gray!6}{100} & \cellcolor{gray!6}{0}\\
$3.HT(1,1.2\tau)$ & 10 & 1158 & 2498 & 1.001 & 1.004 & 0 & 100 & 0\\
$4.HT(4,1)$ & 10 & 2059 & 2819 & 1.001 & 1.003 & 0 & 100 & 0\\
\cellcolor{gray!6}{$5.IG(1/2,\tau^2/2)$} & \cellcolor{gray!6}{10} & \cellcolor{gray!6}{1522} & \cellcolor{gray!6}{2559} & \cellcolor{gray!6}{1.001} & \cellcolor{gray!6}{1.007} & \cellcolor{gray!6}{0} & \cellcolor{gray!6}{100} & \cellcolor{gray!6}{0}\\
\cellcolor{gray!6}{$6.HT(1,\tau)$} & \cellcolor{gray!6}{10} & \cellcolor{gray!6}{1883} & \cellcolor{gray!6}{2624} & \cellcolor{gray!6}{1.001} & \cellcolor{gray!6}{1.003} & \cellcolor{gray!6}{0} & \cellcolor{gray!6}{100} & \cellcolor{gray!6}{0}\\
$7.IG(2,2\tau^2)$ & 10 & 1775 & 2733 & 1.001 & 1.003 & 0 & 100 & 0\\
$8.HT(4,\tau)$ & 10 & 1965 & 2652 & 1.001 & 1.004 & 0 & 100 & 0\\
\cellcolor{gray!6}{$9.IG(5,5\tau^2)$} & \cellcolor{gray!6}{10} & \cellcolor{gray!6}{2227} & \cellcolor{gray!6}{3023} & \cellcolor{gray!6}{1.001} & \cellcolor{gray!6}{1.003} & \cellcolor{gray!6}{0} & \cellcolor{gray!6}{100} & \cellcolor{gray!6}{0}\\
\cellcolor{gray!6}{$10.HT(10,\tau)$} & \cellcolor{gray!6}{10} & \cellcolor{gray!6}{2} & \cellcolor{gray!6}{2679} & \cellcolor{gray!6}{1.016} & \cellcolor{gray!6}{2.443} & \cellcolor{gray!6}{0} & \cellcolor{gray!6}{100} & \cellcolor{gray!6}{0}\\
$11.IG(2,2(c\tau)^2)$ & 10 & 2097 & 2773 & 1.001 & 1.004 & 0 & 100 & 0\\
$12.HT(4,c\tau)$ & 10 & 1814 & 2619 & 1.001 & 1.005 & 0 & 100 & 0\\
\cellcolor{gray!6}{$13.IG(2,2(\tau/c)^2)$} & \cellcolor{gray!6}{10} & \cellcolor{gray!6}{2031} & \cellcolor{gray!6}{2860} & \cellcolor{gray!6}{1.001} & \cellcolor{gray!6}{1.003} & \cellcolor{gray!6}{0} & \cellcolor{gray!6}{100} & \cellcolor{gray!6}{0}\\
\cellcolor{gray!6}{$14.HT(4,\tau/c)$} & \cellcolor{gray!6}{10} & \cellcolor{gray!6}{1650} & \cellcolor{gray!6}{2762} & \cellcolor{gray!6}{1.001} & \cellcolor{gray!6}{1.004} & \cellcolor{gray!6}{0} & \cellcolor{gray!6}{100} & \cellcolor{gray!6}{0}\\
\bottomrule
\end{tabular}
\caption{\label{tab:model2-diag-n10}Diagnostics for Model 2, the longitudinal model, with $n = 10$.}
\end{table}

\begin{table}[tbp]
\centering
\fontsize{10}{12}\selectfont

\begin{tabular}[t]{lrrrrrrrr}
\toprule
Prior & \makecell[c]{True\\Value} & Mean & Median & Bias & \makecell[c]{Rel\\Bias} & RMSE & Cov & \makecell[c]{Interval\\Length}\\
\midrule
\cellcolor{gray!6}{$1.IG(1,1)$} & \cellcolor{gray!6}{0.1} & \cellcolor{gray!6}{0.670} & \cellcolor{gray!6}{0.639} & \cellcolor{gray!6}{0.570} & \cellcolor{gray!6}{5.704} & \cellcolor{gray!6}{0.572} & \cellcolor{gray!6}{0.00} & \cellcolor{gray!6}{0.707}\\
\cellcolor{gray!6}{$2.IG(.001,.001)$} & \cellcolor{gray!6}{0.1} & \cellcolor{gray!6}{0.215} & \cellcolor{gray!6}{0.168} & \cellcolor{gray!6}{0.115} & \cellcolor{gray!6}{1.152} & \cellcolor{gray!6}{0.137} & \cellcolor{gray!6}{1.00} & \cellcolor{gray!6}{0.625}\\
$3.HT(1,1.2\tau)$ & 0.1 & 0.150 & 0.109 & 0.050 & 0.499 & 0.070 & 1.00 & 0.497\\
$4.HT(4,1)$ & 0.1 & 0.309 & 0.277 & 0.209 & 2.090 & 0.229 & 0.98 & 0.763\\
\cellcolor{gray!6}{$5.IG(1/2,\tau^2/2)$} & \cellcolor{gray!6}{0.1} & \cellcolor{gray!6}{0.173} & \cellcolor{gray!6}{0.133} & \cellcolor{gray!6}{0.073} & \cellcolor{gray!6}{0.734} & \cellcolor{gray!6}{0.085} & \cellcolor{gray!6}{1.00} & \cellcolor{gray!6}{0.469}\\
\cellcolor{gray!6}{$6.HT(1,\tau)$} & \cellcolor{gray!6}{0.1} & \cellcolor{gray!6}{0.136} & \cellcolor{gray!6}{0.094} & \cellcolor{gray!6}{0.036} & \cellcolor{gray!6}{0.360} & \cellcolor{gray!6}{0.055} & \cellcolor{gray!6}{1.00} & \cellcolor{gray!6}{0.478}\\
$7.IG(2,2\tau^2)$ & 0.1 & 0.124 & 0.109 & 0.024 & 0.244 & 0.026 & 1.00 & 0.219\\
$8.HT(4,\tau)$ & 0.1 & 0.100 & 0.078 & 0.000 & 0.003 & 0.017 & 1.00 & 0.315\\
\cellcolor{gray!6}{$9.IG(5,5\tau^2)$} & \cellcolor{gray!6}{0.1} & \cellcolor{gray!6}{0.109} & \cellcolor{gray!6}{0.104} & \cellcolor{gray!6}{0.009} & \cellcolor{gray!6}{0.085} & \cellcolor{gray!6}{0.009} & \cellcolor{gray!6}{1.00} & \cellcolor{gray!6}{0.106}\\
\cellcolor{gray!6}{$10.HT(10,\tau)$} & \cellcolor{gray!6}{0.1} & \cellcolor{gray!6}{0.088} & \cellcolor{gray!6}{0.072} & \cellcolor{gray!6}{-0.012} & \cellcolor{gray!6}{-0.116} & \cellcolor{gray!6}{0.016} & \cellcolor{gray!6}{1.00} & \cellcolor{gray!6}{0.254}\\
$11.IG(2,2(c\tau)^2)$ & 0.1 & 0.182 & 0.163 & 0.082 & 0.817 & 0.083 & 0.99 & 0.290\\
$12.HT(4,c\tau)$ & 0.1 & 0.139 & 0.112 & 0.039 & 0.387 & 0.049 & 1.00 & 0.407\\
\cellcolor{gray!6}{$13.IG(2,2(\tau/c)^2)$} & \cellcolor{gray!6}{0.1} & \cellcolor{gray!6}{0.084} & \cellcolor{gray!6}{0.073} & \cellcolor{gray!6}{-0.016} & \cellcolor{gray!6}{-0.164} & \cellcolor{gray!6}{0.017} & \cellcolor{gray!6}{1.00} & \cellcolor{gray!6}{0.153}\\
\cellcolor{gray!6}{$14.HT(4,\tau/c)$} & \cellcolor{gray!6}{0.1} & \cellcolor{gray!6}{0.068} & \cellcolor{gray!6}{0.052} & \cellcolor{gray!6}{-0.032} & \cellcolor{gray!6}{-0.316} & \cellcolor{gray!6}{0.033} & \cellcolor{gray!6}{1.00} & \cellcolor{gray!6}{0.227}\\
\midrule
\cellcolor{gray!6}{$1.IG(1,1)$} & \cellcolor{gray!6}{1.0} & \cellcolor{gray!6}{1.078} & \cellcolor{gray!6}{1.032} & \cellcolor{gray!6}{0.078} & \cellcolor{gray!6}{0.078} & \cellcolor{gray!6}{0.245} & \cellcolor{gray!6}{0.98} & \cellcolor{gray!6}{1.136}\\
\cellcolor{gray!6}{$2.IG(.001,.001)$} & \cellcolor{gray!6}{1.0} & \cellcolor{gray!6}{1.013} & \cellcolor{gray!6}{0.964} & \cellcolor{gray!6}{0.013} & \cellcolor{gray!6}{0.013} & \cellcolor{gray!6}{0.379} & \cellcolor{gray!6}{0.92} & \cellcolor{gray!6}{1.429}\\
$3.HT(1,1.2\tau)$ & 1.0 & 1.049 & 1.002 & 0.049 & 0.049 & 0.329 & 0.94 & 1.351\\
$4.HT(4,1)$ & 1.0 & 1.017 & 0.979 & 0.017 & 0.017 & 0.304 & 0.96 & 1.256\\
\cellcolor{gray!6}{$5.IG(1/2,\tau^2/2)$} & \cellcolor{gray!6}{1.0} & \cellcolor{gray!6}{1.076} & \cellcolor{gray!6}{1.024} & \cellcolor{gray!6}{0.076} & \cellcolor{gray!6}{0.076} & \cellcolor{gray!6}{0.279} & \cellcolor{gray!6}{0.97} & \cellcolor{gray!6}{1.231}\\
\cellcolor{gray!6}{$6.HT(1,\tau)$} & \cellcolor{gray!6}{1.0} & \cellcolor{gray!6}{1.033} & \cellcolor{gray!6}{0.986} & \cellcolor{gray!6}{0.033} & \cellcolor{gray!6}{0.033} & \cellcolor{gray!6}{0.324} & \cellcolor{gray!6}{0.95} & \cellcolor{gray!6}{1.325}\\
$7.IG(2,2\tau^2)$ & 1.0 & 1.072 & 1.034 & 0.072 & 0.072 & 0.199 & 0.99 & 0.997\\
$8.HT(4,\tau)$ & 1.0 & 1.017 & 0.979 & 0.017 & 0.017 & 0.304 & 0.96 & 1.256\\
\cellcolor{gray!6}{$9.IG(5,5\tau^2)$} & \cellcolor{gray!6}{1.0} & \cellcolor{gray!6}{1.053} & \cellcolor{gray!6}{1.028} & \cellcolor{gray!6}{0.053} & \cellcolor{gray!6}{0.053} & \cellcolor{gray!6}{0.129} & \cellcolor{gray!6}{1.00} & \cellcolor{gray!6}{0.762}\\
\cellcolor{gray!6}{$10.HT(10,\tau)$} & \cellcolor{gray!6}{1.0} & \cellcolor{gray!6}{1.009} & \cellcolor{gray!6}{0.974} & \cellcolor{gray!6}{0.009} & \cellcolor{gray!6}{0.009} & \cellcolor{gray!6}{0.294} & \cellcolor{gray!6}{0.96} & \cellcolor{gray!6}{1.213}\\
$11.IG(2,2(c\tau)^2)$ & 1.0 & 1.298 & 1.252 & 0.298 & 0.298 & 0.336 & 0.88 & 1.146\\
$12.HT(4,c\tau)$ & 1.0 & 1.070 & 1.024 & 0.070 & 0.070 & 0.326 & 0.96 & 1.357\\
\cellcolor{gray!6}{$13.IG(2,2(\tau/c)^2)$} & \cellcolor{gray!6}{1.0} & \cellcolor{gray!6}{0.929} & \cellcolor{gray!6}{0.894} & \cellcolor{gray!6}{-0.071} & \cellcolor{gray!6}{-0.071} & \cellcolor{gray!6}{0.228} & \cellcolor{gray!6}{0.96} & \cellcolor{gray!6}{0.930}\\
\cellcolor{gray!6}{$14.HT(4,\tau/c)$} & \cellcolor{gray!6}{1.0} & \cellcolor{gray!6}{0.952} & \cellcolor{gray!6}{0.920} & \cellcolor{gray!6}{-0.048} & \cellcolor{gray!6}{-0.048} & \cellcolor{gray!6}{0.297} & \cellcolor{gray!6}{0.95} & \cellcolor{gray!6}{1.157}\\
\midrule
\cellcolor{gray!6}{$1.IG(1,1)$} & \cellcolor{gray!6}{10} & \cellcolor{gray!6}{9.518} & \cellcolor{gray!6}{9.141} & \cellcolor{gray!6}{-0.482} & \cellcolor{gray!6}{-0.048} & \cellcolor{gray!6}{2.704} & \cellcolor{gray!6}{0.84} & \cellcolor{gray!6}{8.678}\\
\cellcolor{gray!6}{$2.IG(.001,.001)$} & \cellcolor{gray!6}{10} & \cellcolor{gray!6}{10.664} & \cellcolor{gray!6}{10.144} & \cellcolor{gray!6}{0.664} & \cellcolor{gray!6}{0.066} & \cellcolor{gray!6}{3.037} & \cellcolor{gray!6}{0.91} & \cellcolor{gray!6}{10.925}\\
$3.HT(1,1.2\tau)$ & 10 & 10.686 & 10.192 & 0.686 & 0.069 & 2.904 & 0.91 & 10.684\\
$4.HT(4,1)$ & 10 & 8.745 & 8.454 & -1.255 & -0.126 & 2.711 & 0.80 & 7.253\\
\cellcolor{gray!6}{$5.IG(1/2,\tau^2/2)$} & \cellcolor{gray!6}{10} & \cellcolor{gray!6}{10.645} & \cellcolor{gray!6}{10.179} & \cellcolor{gray!6}{0.645} & \cellcolor{gray!6}{0.064} & \cellcolor{gray!6}{2.708} & \cellcolor{gray!6}{0.93} & \cellcolor{gray!6}{10.217}\\
\cellcolor{gray!6}{$6.HT(1,\tau)$} & \cellcolor{gray!6}{10} & \cellcolor{gray!6}{10.574} & \cellcolor{gray!6}{10.098} & \cellcolor{gray!6}{0.574} & \cellcolor{gray!6}{0.057} & \cellcolor{gray!6}{2.858} & \cellcolor{gray!6}{0.91} & \cellcolor{gray!6}{10.391}\\
$7.IG(2,2\tau^2)$ & 10 & 10.554 & 10.201 & 0.554 & 0.055 & 2.055 & 0.98 & 8.735\\
$8.HT(4,\tau)$ & 10 & 10.449 & 10.027 & 0.449 & 0.045 & 2.699 & 0.91 & 9.870\\
\cellcolor{gray!6}{$9.IG(5,5\tau^2)$} & \cellcolor{gray!6}{10} & \cellcolor{gray!6}{10.407} & \cellcolor{gray!6}{10.172} & \cellcolor{gray!6}{0.407} & \cellcolor{gray!6}{0.041} & \cellcolor{gray!6}{1.389} & \cellcolor{gray!6}{0.99} & \cellcolor{gray!6}{6.959}\\
\cellcolor{gray!6}{$10.HT(10,\tau)$} & \cellcolor{gray!6}{10} & \cellcolor{gray!6}{10.362} & \cellcolor{gray!6}{9.982} & \cellcolor{gray!6}{0.362} & \cellcolor{gray!6}{0.036} & \cellcolor{gray!6}{2.605} & \cellcolor{gray!6}{0.92} & \cellcolor{gray!6}{9.661}\\
$11.IG(2,2(c\tau)^2)$ & 10 & 12.450 & 12.039 & 2.450 & 0.245 & 2.964 & 0.90 & 10.245\\
$12.HT(4,c\tau)$ & 10 & 10.827 & 10.341 & 0.827 & 0.083 & 2.939 & 0.92 & 10.691\\
\cellcolor{gray!6}{$13.IG(2,2(\tau/c)^2)$} & \cellcolor{gray!6}{10} & \cellcolor{gray!6}{9.558} & \cellcolor{gray!6}{9.241} & \cellcolor{gray!6}{-0.442} & \cellcolor{gray!6}{-0.044} & \cellcolor{gray!6}{2.231} & \cellcolor{gray!6}{0.90} & \cellcolor{gray!6}{7.896}\\
\cellcolor{gray!6}{$14.HT(4,\tau/c)$} & \cellcolor{gray!6}{10} & \cellcolor{gray!6}{10.020} & \cellcolor{gray!6}{9.647} & \cellcolor{gray!6}{0.020} & \cellcolor{gray!6}{0.002} & \cellcolor{gray!6}{2.515} & \cellcolor{gray!6}{0.91} & \cellcolor{gray!6}{9.024}\\
\bottomrule
\end{tabular}
\caption{\label{tab:model2-estimates-n10} Summary results for $\tau$ under Model 2, the longitudinal model, with $n = 10$.}
\end{table}

For priors that use the true $\tau$ as the scale hyperparameter (priors 5--10), the HT priors generally had smaller bias and a larger interval length than their IG counterparts.  With increasing $\nu$ the bias decreased for both the IG and HT priors, and the interval length decreased but at a faster rate for the IG than the HT priors (Figure~\ref{fig:figure1}, top center).  Consequently with $\nu=1$ and $\nu=4$ the HT priors had a smaller RMSE than their IG counterparts while the situation was reversed with $\nu=10$.

For the IG prior with $\nu=4$, choosing a scale hyperparameter that is too large  (P-11) resulted in a substantial increase in bias, whereas choosing it to be too small (P-13) slightly decreased bias relative to the prior based on the true $\tau$ (P-7).  The HT priors also overestimated $\tau$ when using a large scale hyperparameter (P-12) and underestimated it with a small choice (P-14).  Absolute bias was larger for the IG than the HT when $\tau$ was overstated (P-11 vs P-12), and was smaller for the IG than the HT when $\tau$ was understated (P-13 vs P-14) (Figure~\ref{fig:figure2}, top center).  

{\bf Medium $\tau=1$}: Convergence diagnostics with $\tau=1$ were considerably improved relative to $\tau=0.01$.  For simulations with $n=10$, all HT priors had DTs, affecting at most 12\% of the datasets, while the only IG prior with DTs was the IG(0.001, 0.001). $\hat{R}$ and ESS were considerably improved relative to $\tau=0.1$.  The median ESS was 1.2 to 1.6 times larger for the IG priors compared to their HT counterparts.  Convergence diagnostics were much improved for simulations with $n=30$, where only two priors had a few DTs for a single dataset.

With $n=10$ the only prior to have coverage below 90\% was the IG with overestimated prior scale parameter (P-11).  This prior had 88\% coverage and the largest bias of all 14 priors.  The HT priors had smaller bias than their IG counterparts, while generally the IG priors had a smaller RMSE (except the P-13, P-14 pair), smaller interval length, and slightly higher coverage than their HT counterparts.    With $n=30$, all priors estimated $\tau$ with little bias and good coverage.  In contrast to $n=10$, with $n=30$ the IG priors using the true $\tau$ for the scale hyperparameter (priors 5--10) all had slightly smaller bias than their HT counterparts, while still having slightly smaller RMSE and interval length.

{\bf Large $\tau = 10$}:  With $\tau = 10$ the issue of divergent transitions has entirely disappeared (Table \ref{tab:model2-diag-n10}).  Although the median ESS and mean $\hat{R}$ are reasonable across all priors, the HT(10, $\tau$) prior (P-10) had an extremely small minimum ESS and large maximum $\hat{R}$.  Diagnostics with $n=30$ were of little concern, apart from some relatively small minimum ESS and maximum $\hat{R}$.  

With $n=10$, two of the priors had poor coverage: the IG(1,1) and the HT(4,1) with 84\% and 80\% coverage respectively.  The HT(4,1) prior also badly underestimated $\tau$ by $12.6\%$.   Among the remaining priors 5--14, only the IG with scale hyperparameter that overestimates $\tau$ (p11) was badly biased, with $\tau$ overestimated by $24.5\%$ on average.  Among priors that use the true $\tau$ for a scale hyperparameter (priors 5--10), the HT priors were somewhat less biased, but had somewhat worse coverage, larger interval length, and larger RMSE  than their IG counterparts.  For the smallest $\nu$, coverage, interval length, and RMSE were similar for the IG and HT pair (P-5, P-6), but the disparity in these quantities increased with larger $\nu$.  With $\nu=10$, the IG prior (P-9) is better in all respects except bias than its HT counterpart (P-10).  For simulated data with $n=30$ observations the same patterns held, but differences between the IG and HT pairs were considerably reduced.

A comparison of priors 11-14 shows that the HT priors are again less sensitive to over and underestimation of the scale parameter than their IG counterparts in terms of bias. The IG prior that overestimates $\tau$ (P-11) had the largest absolute bias of all 14 priors, while having similar coverage (90\% versus 92\%),  similar RMSE, and similar interval length compared to its HT counterpart (P-12).

\subsection{Model 3: the multiple outcomes model}

Our primary focus is the standard error of the random effects, $\tau_r$ and $\tau_b$. In both simulation conditions $\tau_r$ was fixed at a true value of $0.7$ whereas we considered two values of $\tau_b$.   ESS for $\tau_r$ was over $12,000$ for all priors (Supplemental Table~\ref{tab:model3-diag-taur}), and although with small $\tau_b$ many priors show non-zero DTs for $\tau_r$, this was likely a function of the difficulty in estimating $\tau_b$ rather than  $\tau_r$.  The latter parameter was estimated with little bias across all 14 priors under both values of $\tau_b$ (Supplemental Table~\ref{tab:model3-estimates-taur}), and coverage, RMSE, and interval lengths for $\tau_r$ were similar across all priors.  Similarly, the parameters $\beta_1$, and $\sigma^2$, were generally well estimated across all priors and both values of $\tau_b$ (not shown).  We therefore focus on estimation of $\tau_b$. 

{\bf Small $\tau_b = 0.04$:} This small value of $\tau_b$ lead to convergence issues for all HT priors but none of the IG priors (Table~\ref{tab:model3-diag-taub}).  All HT priors had DTs for at least 40\% of the datasets.  

Of the four priors suggested by the literature, the IG(1,1) prior, P-1, performed very poorly with a coverage of zero and an extremely large bias (Table~\ref{tab:model3-estimates-taub}).  The IG(.001,.001) (P-2) and the HT(4,1) (P-4) slightly overestimated $\tau_b$, whereas the HT(1, 1.2$\tau$) (P-3) had little bias, despite issues with DTs.  Except for the IG(1,1) prior, all priors had good coverage.  For priors 5--14 the mean $\hat{R}$s were somewhat larger for the HT than the IG priors.  In addition the median ESS was larger for the IG priors than their HT counterparts by a factor of 1.76 (small $\nu$; P-5 versus P-6) to 2.82 (large $\nu$; P-9 versus P-10). 

\begin{table}[!htb]
\centering
\fontsize{10}{12}\selectfont
\begin{tabular}[t]{lrrrrrrrr}
\toprule
Prior & \makecell[c]{True\\Value} & \makecell[c]{Min\\ESS} & \makecell[c]{Med\\ESS} & \makecell[c]{Mean\\Rhat} & \makecell[c]{Max\\Rhat} & \makecell[c]{Mean\\DT} & \makecell[c]{\% 0\\DT} & \makecell[c]{Max\\DT}\\
\midrule
\cellcolor{gray!6}{$1.IG(1,1)$} & \cellcolor{gray!6}{0.04} & \cellcolor{gray!6}{1802} & \cellcolor{gray!6}{5531} & \cellcolor{gray!6}{1.000} & \cellcolor{gray!6}{1.002} & \cellcolor{gray!6}{0.00} & \cellcolor{gray!6}{100} & \cellcolor{gray!6}{0}\\
\cellcolor{gray!6}{$2.IG(.001,.001)$} & \cellcolor{gray!6}{0.04} & \cellcolor{gray!6}{1513} & \cellcolor{gray!6}{5343} & \cellcolor{gray!6}{1.000} & \cellcolor{gray!6}{1.002} & \cellcolor{gray!6}{0.00} & \cellcolor{gray!6}{100} & \cellcolor{gray!6}{0}\\
$3.HT(1,1.2\tau)$ & 0.04 & 362 & 3377 & 1.002 & 1.027 & 9.11 & 51 & 393\\
$4.HT(4,1)$ & 0.04 & 20 & 3106 & 1.002 & 1.109 & 15.68 & 59 & 1110\\
\cellcolor{gray!6}{$5.IG(.5,\tau^2/2)$} & \cellcolor{gray!6}{0.04} & \cellcolor{gray!6}{2847} & \cellcolor{gray!6}{5772} & \cellcolor{gray!6}{1.000} & \cellcolor{gray!6}{1.002} & \cellcolor{gray!6}{0.00} & \cellcolor{gray!6}{100} & \cellcolor{gray!6}{0}\\
\cellcolor{gray!6}{$6.HT(1,\tau)$} & \cellcolor{gray!6}{0.04} & \cellcolor{gray!6}{6} & \cellcolor{gray!6}{3274} & \cellcolor{gray!6}{1.005} & \cellcolor{gray!6}{1.225} & \cellcolor{gray!6}{42.14} & \cellcolor{gray!6}{46} & \cellcolor{gray!6}{2507}\\
$7.IG(2,2\tau^2)$ & 0.04 & 5530 & 8459 & 1.000 & 1.001 & 0.00 & 100 & 0\\
$8.HT(4,\tau)$ & 0.04 & 120 & 3710 & 1.002 & 1.044 & 10.30 & 50 & 466\\
\cellcolor{gray!6}{$9.IG(5,5\tau^2)$} & \cellcolor{gray!6}{0.04} & \cellcolor{gray!6}{8924} & \cellcolor{gray!6}{11290} & \cellcolor{gray!6}{1.000} & \cellcolor{gray!6}{1.001} & \cellcolor{gray!6}{0.00} & \cellcolor{gray!6}{100} & \cellcolor{gray!6}{0}\\
\cellcolor{gray!6}{$10.HT(10,\tau)$} & \cellcolor{gray!6}{0.04} & \cellcolor{gray!6}{6} & \cellcolor{gray!6}{4009} & \cellcolor{gray!6}{1.004} & \cellcolor{gray!6}{1.260} & \cellcolor{gray!6}{31.86} & \cellcolor{gray!6}{49} & \cellcolor{gray!6}{2512}\\
$11.IG(2,2(c\tau)^2)$ & 0.04 & 6226 & 9003 & 1.000 & 1.001 & 0.00 & 100 & 0\\
$12.HT(4,c\tau)$ & 0.04 & 384 & 4195 & 1.001 & 1.009 & 3.14 & 54 & 72\\
\cellcolor{gray!6}{$13.IG(2,2(c/\tau)^2)$} & \cellcolor{gray!6}{0.04} & \cellcolor{gray!6}{4515} & \cellcolor{gray!6}{7010} & \cellcolor{gray!6}{1.000} & \cellcolor{gray!6}{1.001} & \cellcolor{gray!6}{0.00} & \cellcolor{gray!6}{100} & \cellcolor{gray!6}{0}\\
\cellcolor{gray!6}{$14.HT(4,(\tau/c)^2)$} & \cellcolor{gray!6}{0.04} & \cellcolor{gray!6}{98} & \cellcolor{gray!6}{3120} & \cellcolor{gray!6}{1.002} & \cellcolor{gray!6}{1.025} & \cellcolor{gray!6}{10.40} & \cellcolor{gray!6}{47} & \cellcolor{gray!6}{252}\\\hline
\cellcolor{gray!6}{$1.IG(1,1)$} & \cellcolor{gray!6}{0.16} & \cellcolor{gray!6}{3378} & \cellcolor{gray!6}{5768} & \cellcolor{gray!6}{1.000} & \cellcolor{gray!6}{1.002} & \cellcolor{gray!6}{0} & \cellcolor{gray!6}{100} & \cellcolor{gray!6}{0}\\
\cellcolor{gray!6}{$2.IG(.001,.001)$} & \cellcolor{gray!6}{0.16} & \cellcolor{gray!6}{491} & \cellcolor{gray!6}{4397} & \cellcolor{gray!6}{1.001} & \cellcolor{gray!6}{1.009} & \cellcolor{gray!6}{0} & \cellcolor{gray!6}{100} & \cellcolor{gray!6}{0}\\
$3.HT(1,1.2\tau)$ & 0.16 & 491 & 4897 & 1.001 & 1.008 & 0 & 100 & 0\\
$4.HT(4,1)$ & 0.16 & 284 & 3895 & 1.001 & 1.013 & 0 & 100 & 0\\
\cellcolor{gray!6}{$5.IG(.5,\tau^2/2)$} & \cellcolor{gray!6}{0.16} & \cellcolor{gray!6}{326} & \cellcolor{gray!6}{5070} & \cellcolor{gray!6}{1.001} & \cellcolor{gray!6}{1.008} & \cellcolor{gray!6}{0} & \cellcolor{gray!6}{100} & \cellcolor{gray!6}{0}\\
\cellcolor{gray!6}{$6.HT(1,\tau)$} & \cellcolor{gray!6}{0.16} & \cellcolor{gray!6}{973} & \cellcolor{gray!6}{5336} & \cellcolor{gray!6}{1.000} & \cellcolor{gray!6}{1.006} & \cellcolor{gray!6}{0} & \cellcolor{gray!6}{100} & \cellcolor{gray!6}{0}\\
$7.IG(2,2\tau^2)$ & 0.16 & 2425 & 6831 & 1.000 & 1.002 & 0 & 100 & 0\\
$8.HT(4,\tau)$ & 0.16 & 2714 & 6034 & 1.000 & 1.001 & 0 & 100 & 0\\
\cellcolor{gray!6}{$9.IG(5,5\tau^2)$} & \cellcolor{gray!6}{0.16} & \cellcolor{gray!6}{6032} & \cellcolor{gray!6}{9470} & \cellcolor{gray!6}{1.000} & \cellcolor{gray!6}{1.001} & \cellcolor{gray!6}{0} & \cellcolor{gray!6}{100} & \cellcolor{gray!6}{0}\\
\cellcolor{gray!6}{$10.HT(10,\tau)$} & \cellcolor{gray!6}{0.16} & \cellcolor{gray!6}{2809} & \cellcolor{gray!6}{6790} & \cellcolor{gray!6}{1.000} & \cellcolor{gray!6}{1.004} & \cellcolor{gray!6}{0} & \cellcolor{gray!6}{100} & \cellcolor{gray!6}{0}\\
$11.IG(2,2(c\tau)^2)$ & 0.16 & 3468 & 7152 & 1.000 & 1.002 & 0 & 100 & 0\\
$12.HT(4,c\tau)$ & 0.16 & 2358 & 5509 & 1.000 & 1.003 & 0 & 100 & 0\\
\cellcolor{gray!6}{$13.IG(2,2(c/\tau)^2)$} & \cellcolor{gray!6}{0.16} & \cellcolor{gray!6}{3068} & \cellcolor{gray!6}{7638} & \cellcolor{gray!6}{1.000} & \cellcolor{gray!6}{1.002} & \cellcolor{gray!6}{0} & \cellcolor{gray!6}{100} & \cellcolor{gray!6}{0}\\
\cellcolor{gray!6}{$14.HT(4,(\tau/c)^2)$} & \cellcolor{gray!6}{0.16} & \cellcolor{gray!6}{1961} & \cellcolor{gray!6}{6401} & \cellcolor{gray!6}{1.000} & \cellcolor{gray!6}{1.002} & \cellcolor{gray!6}{0} & \cellcolor{gray!6}{100} & \cellcolor{gray!6}{0}\\
\bottomrule
\end{tabular}
\caption{\label{tab:model3-diag-taub}Diagnostics for $\tau_b$ under Model 3, the multiple outcomes model.}
\end{table}

\begin{table}[!htb]
\centering
\fontsize{10}{12}\selectfont
\begin{tabular}[t]{lrrrrrrrr}
\toprule
Prior & \makecell[c]{True\\Value} & Mean & Median & Bias & \makecell[c]{Rel\\Bias} & RMSE & Cov & \makecell[c]{Interval\\Length}\\
\midrule
\cellcolor{gray!6}{$1.IG(1,1)$} & \cellcolor{gray!6}{0.04} & \cellcolor{gray!6}{0.557} & \cellcolor{gray!6}{0.525} & \cellcolor{gray!6}{0.517} & \cellcolor{gray!6}{12.914} & \cellcolor{gray!6}{0.517} & \cellcolor{gray!6}{0.00} & \cellcolor{gray!6}{0.622}\\
\cellcolor{gray!6}{$2.IG(.001,.001)$} & \cellcolor{gray!6}{0.04} & \cellcolor{gray!6}{0.054} & \cellcolor{gray!6}{0.049} & \cellcolor{gray!6}{0.014} & \cellcolor{gray!6}{0.349} & \cellcolor{gray!6}{0.019} & \cellcolor{gray!6}{0.96} & \cellcolor{gray!6}{0.085}\\
$3.HT(1,1.2\tau)$ & 0.04 & 0.042 & 0.039 & 0.002 & 0.046 & 0.017 & 0.95 & 0.076\\
$4.HT(4,1)$ & 0.04 & 0.052 & 0.046 & 0.012 & 0.291 & 0.024 & 0.95 & 0.104\\
\cellcolor{gray!6}{$5.IG(.5,\tau^2/2)$} & \cellcolor{gray!6}{0.04} & \cellcolor{gray!6}{0.045} & \cellcolor{gray!6}{0.042} & \cellcolor{gray!6}{0.005} & \cellcolor{gray!6}{0.132} & \cellcolor{gray!6}{0.013} & \cellcolor{gray!6}{0.98} & \cellcolor{gray!6}{0.068}\\
\cellcolor{gray!6}{$6.HT(1,\tau)$} & \cellcolor{gray!6}{0.04} & \cellcolor{gray!6}{0.041} & \cellcolor{gray!6}{0.037} & \cellcolor{gray!6}{0.001} & \cellcolor{gray!6}{0.014} & \cellcolor{gray!6}{0.017} & \cellcolor{gray!6}{0.97} & \cellcolor{gray!6}{0.074}\\
$7.IG(2,2\tau^2)$ & 0.04 & 0.044 & 0.042 & 0.004 & 0.107 & 0.008 & 1.00 & 0.050\\
$8.HT(4,\tau)$ & 0.04 & 0.039 & 0.037 & -0.001 & -0.019 & 0.015 & 0.97 & 0.067\\
\cellcolor{gray!6}{$9.IG(5,5\tau^2)$} & \cellcolor{gray!6}{0.04} & \cellcolor{gray!6}{0.043} & \cellcolor{gray!6}{0.041} & \cellcolor{gray!6}{0.003} & \cellcolor{gray!6}{0.065} & \cellcolor{gray!6}{0.005} & \cellcolor{gray!6}{1.00} & \cellcolor{gray!6}{0.035}\\
\cellcolor{gray!6}{$10.HT(10,\tau)$} & \cellcolor{gray!6}{0.04} & \cellcolor{gray!6}{0.038} & \cellcolor{gray!6}{0.036} & \cellcolor{gray!6}{-0.002} & \cellcolor{gray!6}{-0.039} & \cellcolor{gray!6}{0.015} & \cellcolor{gray!6}{0.97} & \cellcolor{gray!6}{0.063}\\
$11.IG(2,2(c\tau)^2)$ & 0.04 & 0.056 & 0.054 & 0.016 & 0.408 & 0.017 & 0.93 & 0.059\\
$12.HT(4,c\tau)$ & 0.04 & 0.043 & 0.040 & 0.003 & 0.077 & 0.017 & 0.96 & 0.075\\
\cellcolor{gray!6}{$13.IG(2,2(c/\tau)^2)$} & \cellcolor{gray!6}{0.04} & \cellcolor{gray!6}{0.036} & \cellcolor{gray!6}{0.034} & \cellcolor{gray!6}{-0.004} & \cellcolor{gray!6}{-0.104} & \cellcolor{gray!6}{0.009} & \cellcolor{gray!6}{1.00} & \cellcolor{gray!6}{0.045}\\
\cellcolor{gray!6}{$14.HT(4,(\tau/c)^2)$} & \cellcolor{gray!6}{0.04} & \cellcolor{gray!6}{0.035} & \cellcolor{gray!6}{0.033} & \cellcolor{gray!6}{-0.005} & \cellcolor{gray!6}{-0.128} & \cellcolor{gray!6}{0.015} & \cellcolor{gray!6}{0.94} & \cellcolor{gray!6}{0.059}\\\hline
\cellcolor{gray!6}{$1.IG(1,1)$} & \cellcolor{gray!6}{0.16} & \cellcolor{gray!6}{0.577} & \cellcolor{gray!6}{0.544} & \cellcolor{gray!6}{0.417} & \cellcolor{gray!6}{2.606} & \cellcolor{gray!6}{0.417} & \cellcolor{gray!6}{0.00} & \cellcolor{gray!6}{0.644}\\
\cellcolor{gray!6}{$2.IG(.001,.001)$} & \cellcolor{gray!6}{0.16} & \cellcolor{gray!6}{0.182} & \cellcolor{gray!6}{0.167} & \cellcolor{gray!6}{0.022} & \cellcolor{gray!6}{0.139} & \cellcolor{gray!6}{0.062} & \cellcolor{gray!6}{0.97} & \cellcolor{gray!6}{0.253}\\
$3.HT(1,1.2\tau)$ & 0.16 & 0.180 & 0.167 & 0.020 & 0.125 & 0.056 & 0.97 & 0.235\\
$4.HT(4,1)$ & 0.16 & 0.202 & 0.183 & 0.042 & 0.264 & 0.075 & 0.93 & 0.306\\
\cellcolor{gray!6}{$5.IG(.5,\tau^2/2)$} & \cellcolor{gray!6}{0.16} & \cellcolor{gray!6}{0.179} & \cellcolor{gray!6}{0.167} & \cellcolor{gray!6}{0.019} & \cellcolor{gray!6}{0.119} & \cellcolor{gray!6}{0.051} & \cellcolor{gray!6}{0.97} & \cellcolor{gray!6}{0.223}\\
\cellcolor{gray!6}{$6.HT(1,\tau)$} & \cellcolor{gray!6}{0.16} & \cellcolor{gray!6}{0.177} & \cellcolor{gray!6}{0.164} & \cellcolor{gray!6}{0.017} & \cellcolor{gray!6}{0.105} & \cellcolor{gray!6}{0.054} & \cellcolor{gray!6}{0.97} & \cellcolor{gray!6}{0.228}\\
$7.IG(2,2\tau^2)$ & 0.16 & 0.174 & 0.166 & 0.014 & 0.087 & 0.035 & 0.97 & 0.172\\
$8.HT(4,\tau)$ & 0.16 & 0.172 & 0.162 & 0.012 & 0.076 & 0.048 & 0.97 & 0.205\\
\cellcolor{gray!6}{$9.IG(5,5\tau^2)$} & \cellcolor{gray!6}{0.16} & \cellcolor{gray!6}{0.169} & \cellcolor{gray!6}{0.164} & \cellcolor{gray!6}{0.009} & \cellcolor{gray!6}{0.056} & \cellcolor{gray!6}{0.022} & \cellcolor{gray!6}{0.98} & \cellcolor{gray!6}{0.126}\\
\cellcolor{gray!6}{$10.HT(10,\tau)$} & \cellcolor{gray!6}{0.16} & \cellcolor{gray!6}{0.170} & \cellcolor{gray!6}{0.161} & \cellcolor{gray!6}{0.010} & \cellcolor{gray!6}{0.063} & \cellcolor{gray!6}{0.045} & \cellcolor{gray!6}{0.97} & \cellcolor{gray!6}{0.195}\\
$11.IG(2,2(c\tau)^2)$ & 0.16 & 0.214 & 0.204 & 0.054 & 0.339 & 0.060 & 0.89 & 0.210\\
$12.HT(4,c\tau)$ & 0.16 & 0.182 & 0.170 & 0.022 & 0.141 & 0.055 & 0.96 & 0.233\\
\cellcolor{gray!6}{$13.IG(2,2(c/\tau)^2)$} & \cellcolor{gray!6}{0.16} & \cellcolor{gray!6}{0.151} & \cellcolor{gray!6}{0.145} & \cellcolor{gray!6}{-0.009} & \cellcolor{gray!6}{-0.053} & \cellcolor{gray!6}{0.037} & \cellcolor{gray!6}{0.98} & \cellcolor{gray!6}{0.150}\\
\cellcolor{gray!6}{$14.HT(4,(\tau/c)^2)$} & \cellcolor{gray!6}{0.16} & \cellcolor{gray!6}{0.161} & \cellcolor{gray!6}{0.153} & \cellcolor{gray!6}{0.001} & \cellcolor{gray!6}{0.009} & \cellcolor{gray!6}{0.043} & \cellcolor{gray!6}{0.97} & \cellcolor{gray!6}{0.180}\\
\bottomrule
\end{tabular}
\caption{\label{tab:model3-estimates-taub}Summary results for $\tau_b$ under Model 3, the multiple outcomes model.}
\end{table}

 Among priors 5--10 that use the true $\tau_b$ for the scale hyperparameter, the HT priors (P-6, P-8, P-10) were slightly less biased than their IG counterparts (P-5, P-7, P-9), while having larger interval lengths and larger RMSEs than the IG priors.  With increasing $\nu$, bias decreased more for the IG priors than their HT counterparts, while interval length and RMSE increased more for the HT priors so that with $\nu=10$ the IG prior (P-9) had better overall performance than its HT counterpart (P-10) (Figure~\ref{fig:figure1}, top right). 

When the scale hyperparameter is overestimated, both priors had positive bias, with the IG prior (P-11) showing much larger bias than its HT counterpart (P-12).  Using an underestimate of $\tau_b$ for the scale hyperparameter resulted in a slight negative bias for both the IG (P-13) and HT (P-14) priors.  In both cases the IG priors had smaller interval lengths (Figure~\ref{fig:figure2}, top right) and equal or smaller RMSE than their HT counterparts.

{\bf Large $\tau_b=0.16$}:  Convergence diagnostics were considerably better with the larger $\tau_b$.  No prior had DTs, the mean $\hat{R}$'s were very close to 1.000, and the median ESS was at least 4000 and was similar for the IG and HT priors.  The IG(1,1) prior (P-1) again had very large bias and a coverage of 0.  The other three priors suggested by the literature all performed quite well, with the HT(4,1) (P-4) having the largest bias, interval length and RMSE among these three priors.

Priors that used $\tau_b$ as the scale hyperparameter (priors 5-10) all performed well, with good coverage and small but positive bias.  Differences between the IG and HT pairs were slight, with the greatest difference for the largest $\nu=10$ for which the IG prior (P-9) was slightly better in all respects than its HT counterpart (P-10) (Figure~\ref{fig:figure1}, bottom right).

When the scale hyperparameter overestimates $\tau_b$, the IG prior, P-11, performed less well than the HT counterpart (P-12), with larger bias, and 89\% coverage. With a small prior scale hyperparameter (P-13, P-14), both the IG and HT priors had smaller absolute bias, smaller RMSE, and a shorter interval length than when using an inflated prior scale hyperparameter (P-11, P-12).  Estimates of $\tau_b$ were less sensitive to changes in the prior scale hyperparameters under the HT prior compared to the IG (Figure~\ref{fig:figure2}, bottom right).

\section{Discussion}

Inference for the standard deviation of random effects in hierarchical models is known to be sensitive to the choice of its prior distribution.  Using simulations, we addressed the question of how well $\tau$, the random effects standard deviation, can be estimated under IG and HT priors over a range of hyperparameters.  We considered three different models - the 8-schools model, a longitudinal model with random intercepts, and a simple multiple outcomes model. We compared seven IG priors for $\tau^2$ and seven HT priors for $\tau$.  We argue that  results from IG and HT priors with the same prior degrees of freedom and prior scale hyperparameters can directly be compared, despite their quite different distributions. 

Our simulation results showed that estimation of $\tau$ using typical hyperparameters suggested in the literature was sometimes extremely poor.  When $\tau$ is very small, $\tau^2 \sim IG(1,1)$ was a particularly unsuitable prior, with a coverage of 0\% for all three models. 
Although the other three priors that are commonly suggested in the literature had somewhat better performance, in some situations the IG($0.001, 0.001$) and HT($4,1$) priors estimated $\tau$ with considerable bias.  These results suggest that the expected magnitude of the SD of the random effects in a hierarchical model should be considered when choosing hyperparameters for either an IG or HT prior.  

When choosing an IG prior, many investigators use IG($\epsilon, \epsilon$) for some small value of $\epsilon$.  We agree with \cite{Wakefield:2007, Wakefield:2009} that the two hyperparameters should be chosen separately.   Our simulations showed that choosing a scale hyperparameter for an IG prior that is larger than the true value can cause serious overestimation of $\tau$, as noted by others \citep{Crainiceanu:2008}, whereas choosing a scale hyperparameter that is smaller than $\tau$ by the same factor leads to far smaller absolute bias.  

Except for very small degrees of freedom, $\nu$, when $\tau^2 \sim IG$ the draws of $\tau$ have a smaller SD than draws of $\tau$ from a HT distribution with the same scale and same value of  $\nu$.  As $\nu$ increases, the posterior interval width decreases far more rapidly under the IG then the HT prior.  Therefore the choice of whether to use an IG or HT prior, as well as the choice of their prior hyperparameters, could be viewed as a bias-variance tradeoff.  In particular, because its density is more peaked than the HT,  the IG prior can lead to more bias in the posterior estimate of $\tau$ especially when the scale hyperparameter is not chosen well.  Conversely, the HT prior can lead to larger variance in the estimate of $\tau$ and wider posterior intervals,  which may be due to the strong assumption that $\tau$ can be very close to zero for any scale hyperparameter.

Results for IG and HT priors that use the true value of $\tau$ as the basis for the scale hyperparameter were, in some cases, very sensitive to the prior degrees of freedom, $\nu$.  This sensitivity appears to be related to the information in the likelihood.  Specifically, the  length of the 95\% posterior intervals for the 8-schools model (with $8$ observations from which to estimate $J=8$ random effects) was 14.3 times longer for the IG prior and 5.4 times longer for the HT prior for $\nu=1$ as compared to $\nu=10$.  In comparison, for the multiple outcomes model with $n=4900$ correlated observations with which we estimate $J=7$ random effects, the length of the posterior intervals were far less sensitive to the choice of $\nu$.  Regardless of the model considered or the true $\tau$, for the three values of $\nu$ that we considered here, the interval lengths based on an IG prior were nearly always shorter than those resulting from the paired HT prior.  

In summary, we found that it can be especially challenging to sample from the posterior of $\tau$ when the true value is very small, often leading to divergent transitions in Stan for HT priors but rarely for  IG priors.   When choosing hyperparameters for either an IG or HT prior for the variance or SD of random effects, we argue that it is most important to choose a reasonable scale hyperparameter based on some {\it a priori} knowledge of the magnitude of $\tau$. 
Although it is less clear what to choose for the {\it a priori} degrees of freedom, $\nu$, one should bear in mind that when the number of random effects is small, $\nu$ can have a large impact on the length of the posterior interval for $\tau$.  Finally, when it is possible to choose the prior scale hyperparameter well, the choice of whether to use an IG or HT prior may depend on the magnitude of $\tau$ and whether one is primarily interested in it's point estimate or interval length.   
When $\tau$ is small, a HT prior is likely to be preferred when the point estimate of $\tau$ is of most interest (although convergence may be an issue), whereas an IG prior may be preferred based on interval length.  For large $\tau,$  the IG prior may be preferable based on interval length, and can sometimes result in smaller bias than the HT, especially when the data has few random effects or when $\nu$ can be chosen to be relatively large.   However the consequences of choosing a poor prior scale hyperparameter are more severe for the IG than the HT prior, especially if this hyperparameter is chosen to be too large. More work is needed to determine whether these general conclusions hold for $\nu$ outside the range we considered here, and to what extent varying $\nu$ affects posterior inference when the prior scale hyperparameter for $\tau$ is not chosen well.

\clearpage

\bibliographystyle{chicago}
\bibliography{IGHTpriors}



\vspace{1in}
{\bf Acknowledgements:}  Research reported in this publication was supported by the  National Institutes of Health (NIH) under award numbers P30ES001247, T32ES007271, R01HL137811, T32GM007356, and 3UH2AR067690.  The content is solely the responsibility of the authors and does not necessarily represent the official views of the NIH.

\clearpage
\setcounter{page}{1}
\begin{center}
  \section*{Supplemental materials for}
  \end{center}

\label{supp:model2}

\begin{center}
  {\large \bf Selection of inverse gamma and half-t priors for hierarchical models: sensitivity and recommendations}\\[2ex]
  \normalsize{Zachary Brehm,
  Aaron Wagner,
Erik VonKaenel,
David Burton,
Samuel J. Weisenthal,
Martin Cole,
Yiping Pang,
and Sally W. Thurston}\\[1ex]
\end{center}
\renewcommand{\thefootnote}{\fnsymbol{footnote}}


\begin{table}[bp!]
\centering
\fontsize{10}{12}\selectfont
\begin{tabular}[t]{lrrrrr}
\toprule
Condition & Mean & Rel Bias & 2.5\% & 97.5\% & $\% 0 \, \hat{\tau}$\\
\midrule
\cellcolor{gray!6}{$n = 10, \, \tau = 0.1$} & \cellcolor{gray!6}{0.145} & \cellcolor{gray!6}{0.449} & \cellcolor{gray!6}{0.000} & \cellcolor{gray!6}{0.534} & \cellcolor{gray!6}{0.51}\\
$n = 10, \, \tau = 1$ & 0.983 & -0.017 & 0.467 & 1.578 & 0.00\\
\cellcolor{gray!6}{$n = 10, \, \tau = 10$} & \cellcolor{gray!6}{9.836} & \cellcolor{gray!6}{-0.016} & \cellcolor{gray!6}{4.652} & \cellcolor{gray!6}{14.908} & \cellcolor{gray!6}{0.00}\\
$n = 30, \, \tau = 0.1$ & 0.148 & 0.476 & 0.000 & 0.426 & 0.37\\
\cellcolor{gray!6}{$n = 30, \, \tau = 1$} & \cellcolor{gray!6}{0.970} & \cellcolor{gray!6}{-0.030} &  \cellcolor{gray!6}{0.696} & \cellcolor{gray!6}{1.257} & \cellcolor{gray!6}{0.00}\\
$n = 30, \, \tau = 10$ & 9.829 & -0.017 & 7.644 & 12.094 & 0.00\\
\bottomrule
\end{tabular}
\caption{\label{tab:lme4}Estimates of $\hat{\tau}$ from lme4 for Model 2, the longitudinal model.}
\end{table}


\begin{table}
\centering
\fontsize{10}{12}\selectfont
\begin{tabular}[t]{lrrrrrrrr}
\toprule
Prior & \makecell[c]{True\\Value} & \makecell[c]{Min\\ESS} & \makecell[c]{Med\\ESS} & \makecell[c]{Mean\\Rhat} & \makecell[c]{Max\\Rhat} & \makecell[c]{Mean\\Div} & \makecell[c]{\% 0\\DT} & \makecell[c]{Max\\DT}\\
\midrule
\cellcolor{gray!6}{$1.IG(1,1)$} & \cellcolor{gray!6}{0.1} & \cellcolor{gray!6}{4149} & \cellcolor{gray!6}{5939} & \cellcolor{gray!6}{1.000} & \cellcolor{gray!6}{1.001} & \cellcolor{gray!6}{0.00} & \cellcolor{gray!6}{100} & \cellcolor{gray!6}{0}\\
\cellcolor{gray!6}{$2.IG(.001,.001)$} & \cellcolor{gray!6}{0.1} & \cellcolor{gray!6}{108} & \cellcolor{gray!6}{218} & \cellcolor{gray!6}{1.018} & \cellcolor{gray!6}{1.046} & \cellcolor{gray!6}{0.00} & \cellcolor{gray!6}{100} & \cellcolor{gray!6}{0}\\
$3.HT(1,1.2\tau)$ & 0.1 & 28 & 139 & 1.033 & 1.151 & 1.31 & 74 & 56\\
$4.HT(4,1)$ & 0.1 & 51 & 183 & 1.023 & 1.075 & 0.28 & 87 & 4\\
\cellcolor{gray!6}{$5.IG(1/2,\tau^2/2)$} & \cellcolor{gray!6}{0.1} & \cellcolor{gray!6}{169} & \cellcolor{gray!6}{369} & \cellcolor{gray!6}{1.010} & \cellcolor{gray!6}{1.041} & \cellcolor{gray!6}{0.00} & \cellcolor{gray!6}{100} & \cellcolor{gray!6}{0}\\
\cellcolor{gray!6}{$6.HT(1,\tau)$} & \cellcolor{gray!6}{0.1} & \cellcolor{gray!6}{38} & \cellcolor{gray!6}{137} & \cellcolor{gray!6}{1.031} & \cellcolor{gray!6}{1.130} & \cellcolor{gray!6}{5.96} & \cellcolor{gray!6}{69} & \cellcolor{gray!6}{312}\\
$7.IG(2,2\tau^2)$ & 0.1 & 339 & 743 & 1.005 & 1.022 & 0.00 & 100 & 0\\
$8.HT(4,\tau)$ & 0.1 & 53 & 159 & 1.024 & 1.102 & 1.38 & 74 & 88\\
\cellcolor{gray!6}{$9.IG(5,5\tau^2)$} & \cellcolor{gray!6}{0.1} & \cellcolor{gray!6}{855} & \cellcolor{gray!6}{1701} & \cellcolor{gray!6}{1.002} & \cellcolor{gray!6}{1.009} & \cellcolor{gray!6}{0.00} & \cellcolor{gray!6}{100} & \cellcolor{gray!6}{0}\\
\cellcolor{gray!6}{$10.HT(10,\tau)$} & \cellcolor{gray!6}{0.1} & \cellcolor{gray!6}{45} & \cellcolor{gray!6}{150} & \cellcolor{gray!6}{1.030} & \cellcolor{gray!6}{1.108} & \cellcolor{gray!6}{0.59} & \cellcolor{gray!6}{79} & \cellcolor{gray!6}{12}\\
$11.IG(2,2(c\tau)^2)$ & 0.1 & 436 & 1026 & 1.004 & 1.016 & 0.00 & 100 & 0\\
$12.HT(4,c\tau)$ & 0.1 & 28 & 155 & 1.029 & 1.105 & 1.35 & 76 & 71\\
\cellcolor{gray!6}{$13.IG(2,2(\tau/c)^2)$} & \cellcolor{gray!6}{0.1} & \cellcolor{gray!6}{208} & \cellcolor{gray!6}{622} & \cellcolor{gray!6}{1.007} & \cellcolor{gray!6}{1.027} & \cellcolor{gray!6}{0.00} & \cellcolor{gray!6}{100} & \cellcolor{gray!6}{0}\\
\cellcolor{gray!6}{$14.HT(4,\tau/c)$} & \cellcolor{gray!6}{0.1} & \cellcolor{gray!6}{38} & \cellcolor{gray!6}{147} & \cellcolor{gray!6}{1.030} & \cellcolor{gray!6}{1.109} & \cellcolor{gray!6}{2.59} & \cellcolor{gray!6}{74} & \cellcolor{gray!6}{110}\\
\midrule
\cellcolor{gray!6}{$1.IG(1,1)$} & \cellcolor{gray!6}{1} & \cellcolor{gray!6}{5498} & \cellcolor{gray!6}{7798} & \cellcolor{gray!6}{1.000} & \cellcolor{gray!6}{1.001} & \cellcolor{gray!6}{0.00} & \cellcolor{gray!6}{100} & \cellcolor{gray!6}{0}\\
\cellcolor{gray!6}{$2.IG(.001,.001)$} & \cellcolor{gray!6}{1} & \cellcolor{gray!6}{149} & \cellcolor{gray!6}{6978} & \cellcolor{gray!6}{1.001} & \cellcolor{gray!6}{1.025} & \cellcolor{gray!6}{0.00} & \cellcolor{gray!6}{100} & \cellcolor{gray!6}{0}\\
$3.HT(1,1.2\tau)$ & 1 & 169 & 7193 & 1.001 & 1.033 & 0.00 & 100 & 0\\
$4.HT(4,1)$ & 1 & 128 & 7043 & 1.000 & 1.023 & 0.23 & 99 & 23\\
\cellcolor{gray!6}{$5.IG(1/2,\tau^2/2)$} & \cellcolor{gray!6}{1} & \cellcolor{gray!6}{4270} & \cellcolor{gray!6}{7475} & \cellcolor{gray!6}{1.000} & \cellcolor{gray!6}{1.001} & \cellcolor{gray!6}{0.00} & \cellcolor{gray!6}{100} & \cellcolor{gray!6}{0}\\
\cellcolor{gray!6}{$6.HT(1,\tau)$} & \cellcolor{gray!6}{1} & \cellcolor{gray!6}{169} & \cellcolor{gray!6}{7357} & \cellcolor{gray!6}{1.000} & \cellcolor{gray!6}{1.008} & \cellcolor{gray!6}{0.00} & \cellcolor{gray!6}{100} & \cellcolor{gray!6}{0}\\
$7.IG(2,2\tau^2)$ & 1 & 6061 & 8332 & 1.000 & 1.001 & 0.00 & 100 & 0\\
$8.HT(4,\tau)$ & 1 & 128 & 7043 & 1.000 & 1.023 & 0.23 & 99 & 23\\
\cellcolor{gray!6}{$9.IG(5,5\tau^2)$} & \cellcolor{gray!6}{1} & \cellcolor{gray!6}{6270} & \cellcolor{gray!6}{9459} & \cellcolor{gray!6}{1.000} & \cellcolor{gray!6}{1.001} & \cellcolor{gray!6}{0.00} & \cellcolor{gray!6}{100} & \cellcolor{gray!6}{0}\\
\cellcolor{gray!6}{$10.HT(10,\tau)$} & \cellcolor{gray!6}{1} & \cellcolor{gray!6}{245} & \cellcolor{gray!6}{7447} & \cellcolor{gray!6}{1.001} & \cellcolor{gray!6}{1.026} & \cellcolor{gray!6}{0.00} & \cellcolor{gray!6}{100} & \cellcolor{gray!6}{0}\\
$11.IG(2,2(c\tau)^2)$ & 1 & 6149 & 8951 & 1.000 & 1.001 & 0.00 & 100 & 0\\
$12.HT(4,c\tau)$ & 1 & 155 & 7441 & 1.000 & 1.027 & 0.00 & 100 & 0\\
\cellcolor{gray!6}{$13.IG(2,2(\tau/c)^2)$} & \cellcolor{gray!6}{1} & \cellcolor{gray!6}{4683} & \cellcolor{gray!6}{7417} & \cellcolor{gray!6}{1.000} & \cellcolor{gray!6}{1.002} & \cellcolor{gray!6}{0.00} & \cellcolor{gray!6}{100} & \cellcolor{gray!6}{0}\\
\cellcolor{gray!6}{$14.HT(4,\tau/c)$} & \cellcolor{gray!6}{1} & \cellcolor{gray!6}{167} & \cellcolor{gray!6}{7167} & \cellcolor{gray!6}{1.000} & \cellcolor{gray!6}{1.026} & \cellcolor{gray!6}{0.00} & \cellcolor{gray!6}{100} & \cellcolor{gray!6}{0}\\
\midrule
\cellcolor{gray!6}{$1.IG(1,1)$} & \cellcolor{gray!6}{10} & \cellcolor{gray!6}{865} & \cellcolor{gray!6}{2405} & \cellcolor{gray!6}{1.002} & \cellcolor{gray!6}{1.012} & \cellcolor{gray!6}{0} & \cellcolor{gray!6}{100} & \cellcolor{gray!6}{0}\\
\cellcolor{gray!6}{$2.IG(.001,.001)$} & \cellcolor{gray!6}{10} & \cellcolor{gray!6}{106} & \cellcolor{gray!6}{2185} & \cellcolor{gray!6}{1.002} & \cellcolor{gray!6}{1.032} & \cellcolor{gray!6}{0} & \cellcolor{gray!6}{100} & \cellcolor{gray!6}{0}\\
$3.HT(1,1.2\tau)$ & 10 & 853 & 2315 & 1.002 & 1.007 & 0 & 100 & 0\\
$4.HT(4,1)$ & 10 & 1139 & 2465 & 1.002 & 1.012 & 0 & 100 & 0\\
\cellcolor{gray!6}{$5.IG(1/2,\tau^2/2)$} & \cellcolor{gray!6}{10} & \cellcolor{gray!6}{1066} & \cellcolor{gray!6}{2443} & \cellcolor{gray!6}{1.002} & \cellcolor{gray!6}{1.007} & \cellcolor{gray!6}{0} & \cellcolor{gray!6}{100} & \cellcolor{gray!6}{0}\\
\cellcolor{gray!6}{$6.HT(1,\tau)$} & \cellcolor{gray!6}{10} & \cellcolor{gray!6}{1026} & \cellcolor{gray!6}{2278} & \cellcolor{gray!6}{1.002} & \cellcolor{gray!6}{1.009} & \cellcolor{gray!6}{0} & \cellcolor{gray!6}{100} & \cellcolor{gray!6}{0}\\
$7.IG(2,2\tau^2)$ & 10 & 1061 & 2296 & 1.002 & 1.004 & 0 & 100 & 0\\
$8.HT(4,\tau)$ & 10 & 376 & 2346 & 1.002 & 1.013 & 0 & 100 & 0\\
\cellcolor{gray!6}{$9.IG(5,5\tau^2)$} & \cellcolor{gray!6}{10} & \cellcolor{gray!6}{960} & \cellcolor{gray!6}{2527} & \cellcolor{gray!6}{1.002} & \cellcolor{gray!6}{1.009} & \cellcolor{gray!6}{0} & \cellcolor{gray!6}{100} & \cellcolor{gray!6}{0}\\
\cellcolor{gray!6}{$10.HT(10,\tau)$} & \cellcolor{gray!6}{10} & \cellcolor{gray!6}{958} & \cellcolor{gray!6}{2269} & \cellcolor{gray!6}{1.002} & \cellcolor{gray!6}{1.006} & \cellcolor{gray!6}{0} & \cellcolor{gray!6}{100} & \cellcolor{gray!6}{0}\\
$11.IG(2,2(c\tau)^2)$ & 10 & 938 & 2432 & 1.002 & 1.007 & 0 & 100 & 0\\
$12.HT(4,c\tau)$ & 10 & 1277 & 2437 & 1.002 & 1.008 & 0 & 100 & 0\\
\cellcolor{gray!6}{$13.IG(2,2(\tau/c)^2)$} & \cellcolor{gray!6}{10} & \cellcolor{gray!6}{703} & \cellcolor{gray!6}{2469} & \cellcolor{gray!6}{1.002} & \cellcolor{gray!6}{1.007} & \cellcolor{gray!6}{0} & \cellcolor{gray!6}{100} & \cellcolor{gray!6}{0}\\
\cellcolor{gray!6}{$14.HT(4,\tau/c)$} & \cellcolor{gray!6}{10} & \cellcolor{gray!6}{751} & \cellcolor{gray!6}{2414} & \cellcolor{gray!6}{1.002} & \cellcolor{gray!6}{1.018} & \cellcolor{gray!6}{0} & \cellcolor{gray!6}{100} & \cellcolor{gray!6}{0}\\
\bottomrule
\end{tabular}
\caption{\label{tab:model2-diag-n30}Diagnostics for $\tau$ under Model 2, the longitudinal model, with $n = 30$.}
\end{table}


\begin{table}
\centering
\fontsize{10}{12}\selectfont
\begin{tabular}[t]{lrrrrrrrr}
\toprule
Prior & \makecell[c]{True\\Value} & Mean & Median & Bias & \makecell[c]{Rel\\Bias} & RMSE & Cov & \makecell[c]{Interval\\Length}\\
\midrule
\cellcolor{gray!6}{$1.IG(1,1)$} & \cellcolor{gray!6}{0.1} & \cellcolor{gray!6}{0.506} & \cellcolor{gray!6}{0.496} & \cellcolor{gray!6}{0.406} & \cellcolor{gray!6}{4.059} & \cellcolor{gray!6}{0.407} & \cellcolor{gray!6}{0.00} & \cellcolor{gray!6}{0.355}\\
\cellcolor{gray!6}{$2.IG(.001,.001)$} & \cellcolor{gray!6}{0.1} & \cellcolor{gray!6}{0.174} & \cellcolor{gray!6}{0.151} & \cellcolor{gray!6}{0.074} & \cellcolor{gray!6}{0.737} & \cellcolor{gray!6}{0.099} & \cellcolor{gray!6}{1.00} & \cellcolor{gray!6}{0.405}\\
$3.HT(1,1.2\tau)$ & 0.1 & 0.136 & 0.113 & 0.036 & 0.361 & 0.064 & 1.00 & 0.370\\
$4.HT(4,1)$ & 0.1 & 0.225 & 0.214 & 0.125 & 1.249 & 0.146 & 0.98 & 0.465\\
\cellcolor{gray!6}{$5.IG(1/2,\tau^2/2)$} & \cellcolor{gray!6}{0.1} & \cellcolor{gray!6}{0.156} & \cellcolor{gray!6}{0.133} & \cellcolor{gray!6}{0.056} & \cellcolor{gray!6}{0.564} & \cellcolor{gray!6}{0.072} & \cellcolor{gray!6}{1.00} & \cellcolor{gray!6}{0.342}\\
\cellcolor{gray!6}{$6.HT(1,\tau)$} & \cellcolor{gray!6}{0.1} & \cellcolor{gray!6}{0.125} & \cellcolor{gray!6}{0.099} & \cellcolor{gray!6}{0.025} & \cellcolor{gray!6}{0.249} & \cellcolor{gray!6}{0.056} & \cellcolor{gray!6}{1.00} & \cellcolor{gray!6}{0.359}\\
$7.IG(2,2\tau^2)$ & 0.1 & 0.125 & 0.111 & 0.025 & 0.255 & 0.029 & 1.00 & 0.211\\
$8.HT(4,\tau)$ & 0.1 & 0.103 & 0.084 & 0.003 & 0.031 & 0.027 & 1.00 & 0.292\\
\cellcolor{gray!6}{$9.IG(5,5\tau^2)$} & \cellcolor{gray!6}{0.1} & \cellcolor{gray!6}{0.109} & \cellcolor{gray!6}{0.104} & \cellcolor{gray!6}{0.009} & \cellcolor{gray!6}{0.089} & \cellcolor{gray!6}{0.009} & \cellcolor{gray!6}{1.00} & \cellcolor{gray!6}{0.107}\\
\cellcolor{gray!6}{$10.HT(10,\tau)$} & \cellcolor{gray!6}{0.1} & \cellcolor{gray!6}{0.092} & \cellcolor{gray!6}{0.077} & \cellcolor{gray!6}{-0.008} & \cellcolor{gray!6}{-0.076} & \cellcolor{gray!6}{0.019} & \cellcolor{gray!6}{1.00} & \cellcolor{gray!6}{0.251}\\
$11.IG(2,2(c\tau)^2)$ & 0.1 & 0.178 & 0.164 & 0.078 & 0.781 & 0.081 & 0.99 & 0.252\\
$12.HT(4,c\tau)$ & 0.1 & 0.139 & 0.121 & 0.039 & 0.385 & 0.056 & 1.00 & 0.347\\
\cellcolor{gray!6}{$13.IG(2,2(\tau/c)^2)$} & \cellcolor{gray!6}{0.1} & \cellcolor{gray!6}{0.085} & \cellcolor{gray!6}{0.074} & \cellcolor{gray!6}{-0.015} & \cellcolor{gray!6}{-0.154} & \cellcolor{gray!6}{0.017} & \cellcolor{gray!6}{1.00} & \cellcolor{gray!6}{0.158}\\
\cellcolor{gray!6}{$14.HT(4,\tau/c)$} & \cellcolor{gray!6}{0.1} & \cellcolor{gray!6}{0.072} & \cellcolor{gray!6}{0.055} & \cellcolor{gray!6}{-0.028} & \cellcolor{gray!6}{-0.282} & \cellcolor{gray!6}{0.032} & \cellcolor{gray!6}{1.00} & \cellcolor{gray!6}{0.229}\\
\midrule
\cellcolor{gray!6}{$1.IG(1,1)$} & \cellcolor{gray!6}{1.0} & \cellcolor{gray!6}{0.996} & \cellcolor{gray!6}{0.982} & \cellcolor{gray!6}{-0.004} & \cellcolor{gray!6}{-0.004} & \cellcolor{gray!6}{0.147} & \cellcolor{gray!6}{0.97} & \cellcolor{gray!6}{0.633}\\
\cellcolor{gray!6}{$2.IG(.001,.001)$} & \cellcolor{gray!6}{1.0} & \cellcolor{gray!6}{0.985} & \cellcolor{gray!6}{0.970} & \cellcolor{gray!6}{-0.015} & \cellcolor{gray!6}{-0.015} & \cellcolor{gray!6}{0.180} & \cellcolor{gray!6}{0.96} & \cellcolor{gray!6}{0.685}\\
$3.HT(1,1.2\tau)$ & 1.0 & 0.991 & 0.977 & -0.009 & -0.009 & 0.172 & 0.96 & 0.671\\
$4.HT(4,1)$ & 1.0 & 0.985 & 0.972 & -0.015 & -0.015 & 0.169 & 0.96 & 0.661\\
\cellcolor{gray!6}{$5.IG(1/2,\tau^2/2)$} & \cellcolor{gray!6}{1.0} & \cellcolor{gray!6}{0.992} & \cellcolor{gray!6}{0.978} & \cellcolor{gray!6}{-0.008} & \cellcolor{gray!6}{-0.008} & \cellcolor{gray!6}{0.157} & \cellcolor{gray!6}{0.97} & \cellcolor{gray!6}{0.654}\\
\cellcolor{gray!6}{$6.HT(1,\tau)$} & \cellcolor{gray!6}{1.0} & \cellcolor{gray!6}{0.987} & \cellcolor{gray!6}{0.973} & \cellcolor{gray!6}{-0.013} & \cellcolor{gray!6}{-0.013} & \cellcolor{gray!6}{0.171} & \cellcolor{gray!6}{0.96} & \cellcolor{gray!6}{0.668}\\
$7.IG(2,2\tau^2)$ & 1.0 & 1.001 & 0.987 & 0.001 & 0.001 & 0.132 & 0.97 & 0.602\\
$8.HT(4,\tau)$ & 1.0 & 0.985 & 0.972 & -0.015 & -0.015 & 0.169 & 0.96 & 0.661\\
\cellcolor{gray!6}{$9.IG(5,5\tau^2)$} & \cellcolor{gray!6}{1.0} & \cellcolor{gray!6}{1.006} & \cellcolor{gray!6}{0.994} & \cellcolor{gray!6}{0.006} & \cellcolor{gray!6}{0.006} & \cellcolor{gray!6}{0.103} & \cellcolor{gray!6}{0.98} & \cellcolor{gray!6}{0.532}\\
\cellcolor{gray!6}{$10.HT(10,\tau)$} & \cellcolor{gray!6}{1.0} & \cellcolor{gray!6}{0.985} & \cellcolor{gray!6}{0.971} & \cellcolor{gray!6}{-0.015} & \cellcolor{gray!6}{-0.015} & \cellcolor{gray!6}{0.167} & \cellcolor{gray!6}{0.96} & \cellcolor{gray!6}{0.656}\\
$11.IG(2,2(c\tau)^2)$ & 1.0 & 1.109 & 1.094 & 0.109 & 0.109 & 0.159 & 0.96 & 0.629\\
$12.HT(4,c\tau)$ & 1.0 & 1.000 & 0.985 & 0.000 & 0.000 & 0.172 & 0.97 & 0.677\\
\cellcolor{gray!6}{$13.IG(2,2(\tau/c)^2)$} & \cellcolor{gray!6}{1.0} & \cellcolor{gray!6}{0.939} & \cellcolor{gray!6}{0.927} & \cellcolor{gray!6}{-0.061} & \cellcolor{gray!6}{-0.061} & \cellcolor{gray!6}{0.158} & \cellcolor{gray!6}{0.94} & \cellcolor{gray!6}{0.598}\\
\cellcolor{gray!6}{$14.HT(4,\tau/c)$} & \cellcolor{gray!6}{1.0} & \cellcolor{gray!6}{0.964} & \cellcolor{gray!6}{0.951} & \cellcolor{gray!6}{-0.036} & \cellcolor{gray!6}{-0.036} & \cellcolor{gray!6}{0.168} & \cellcolor{gray!6}{0.95} & \cellcolor{gray!6}{0.641}\\
\midrule
\cellcolor{gray!6}{$1.IG(1,1)$} & \cellcolor{gray!6}{10} & \cellcolor{gray!6}{9.741} & \cellcolor{gray!6}{9.610} & \cellcolor{gray!6}{-0.259} & \cellcolor{gray!6}{-0.026} & \cellcolor{gray!6}{1.251} & \cellcolor{gray!6}{0.94} & \cellcolor{gray!6}{5.007}\\
\cellcolor{gray!6}{$2.IG(.001,.001)$} & \cellcolor{gray!6}{10} & \cellcolor{gray!6}{10.089} & \cellcolor{gray!6}{9.942} & \cellcolor{gray!6}{0.089} & \cellcolor{gray!6}{0.009} & \cellcolor{gray!6}{1.265} & \cellcolor{gray!6}{0.97} & \cellcolor{gray!6}{5.393}\\
$3.HT(1,1.2\tau)$ & 10 & 10.112 & 9.969 & 0.112 & 0.011 & 1.240 & 0.97 & 5.366\\
$4.HT(4,1)$ & 10 & 9.451 & 9.329 & -0.549 & -0.055 & 1.296 & 0.93 & 4.720\\
\cellcolor{gray!6}{$5.IG(1/2,\tau^2/2)$} & \cellcolor{gray!6}{10} & \cellcolor{gray!6}{10.082} & \cellcolor{gray!6}{9.940} & \cellcolor{gray!6}{0.082} & \cellcolor{gray!6}{0.008} & \cellcolor{gray!6}{1.219} & \cellcolor{gray!6}{0.97} & \cellcolor{gray!6}{5.265}\\
\cellcolor{gray!6}{$6.HT(1,\tau)$} & \cellcolor{gray!6}{10} & \cellcolor{gray!6}{10.073} & \cellcolor{gray!6}{9.929} & \cellcolor{gray!6}{0.073} & \cellcolor{gray!6}{0.007} & \cellcolor{gray!6}{1.237} & \cellcolor{gray!6}{0.97} & \cellcolor{gray!6}{5.324}\\
$7.IG(2,2\tau^2)$ & 10 & 10.084 & 9.953 & 0.084 & 0.008 & 1.104 & 0.98 & 5.029\\
$8.HT(4,\tau)$ & 10 & 10.077 & 9.934 & 0.077 & 0.008 & 1.224 & 0.97 & 5.298\\
\cellcolor{gray!6}{$9.IG(5,5\tau^2)$} & \cellcolor{gray!6}{10} & \cellcolor{gray!6}{10.077} & \cellcolor{gray!6}{9.967} & \cellcolor{gray!6}{0.077} & \cellcolor{gray!6}{0.008} & \cellcolor{gray!6}{0.922} & \cellcolor{gray!6}{0.99} & \cellcolor{gray!6}{4.591}\\
\cellcolor{gray!6}{$10.HT(10,\tau)$} & \cellcolor{gray!6}{10} & \cellcolor{gray!6}{10.061} & \cellcolor{gray!6}{9.922} & \cellcolor{gray!6}{0.061} & \cellcolor{gray!6}{0.006} & \cellcolor{gray!6}{1.218} & \cellcolor{gray!6}{0.97} & \cellcolor{gray!6}{5.253}\\
$11.IG(2,2(c\tau)^2)$ & 10 & 10.844 & 10.705 & 0.844 & 0.084 & 1.324 & 0.96 & 5.390\\
$12.HT(4,c\tau)$ & 10 & 10.161 & 10.015 & 0.161 & 0.016 & 1.260 & 0.97 & 5.398\\
\cellcolor{gray!6}{$13.IG(2,2(\tau/c)^2)$} & \cellcolor{gray!6}{10} & \cellcolor{gray!6}{9.723} & \cellcolor{gray!6}{9.597} & \cellcolor{gray!6}{-0.277} & \cellcolor{gray!6}{-0.028} & \cellcolor{gray!6}{1.175} & \cellcolor{gray!6}{0.95} & \cellcolor{gray!6}{4.838}\\
\cellcolor{gray!6}{$14.HT(4,\tau/c)$} & \cellcolor{gray!6}{10} & \cellcolor{gray!6}{9.933} & \cellcolor{gray!6}{9.798} & \cellcolor{gray!6}{-0.067} & \cellcolor{gray!6}{-0.007} & \cellcolor{gray!6}{1.192} & \cellcolor{gray!6}{0.97} & \cellcolor{gray!6}{5.120}\\
\bottomrule
\end{tabular}
\caption{\label{tab:model2-estimates-n30}Summary results for Model 2, the longitudinal model, with $n = 30$.}
\centering
\end{table}

\clearpage
\label{supp:model3}

\begin{table}[!bh]
\centering
\fontsize{10}{12}\selectfont
\begin{tabular}[t]{lrrrrrrrr}
\toprule
Prior & \makecell[c]{True\\Value} & \makecell[c]{Min\\ESS} & \makecell[c]{Med\\ESS} & \makecell[c]{Mean\\Rhat} & \makecell[c]{Max\\Rhat} & \makecell[c]{Mean\\DT} & \makecell[c]{\% 0\\DT} & \makecell[c]{Max\\DT}\\
\midrule
\cellcolor{gray!6}{$1.IG(1,1)$} & \cellcolor{gray!6}{0.7} & \cellcolor{gray!6}{8402} & \cellcolor{gray!6}{12028} & \cellcolor{gray!6}{1.000} & \cellcolor{gray!6}{1.001} & \cellcolor{gray!6}{0.00} & \cellcolor{gray!6}{100} & \cellcolor{gray!6}{0}\\
\cellcolor{gray!6}{$2.IG(.001,.001)$} & \cellcolor{gray!6}{0.7} & \cellcolor{gray!6}{9657} & \cellcolor{gray!6}{14372} & \cellcolor{gray!6}{1.000} & \cellcolor{gray!6}{1.000} & \cellcolor{gray!6}{0.00} & \cellcolor{gray!6}{100} & \cellcolor{gray!6}{0}\\
$3.HT(1,1.2\tau)$ & 0.7 & 120 & 13487 & 1.000 & 1.030 & 9.11 & 51 & 393\\
$4.HT(4,1)$ & 0.7 & 212 & 13252 & 1.000 & 1.020 & 15.68 & 59 & 1110\\
\cellcolor{gray!6}{$5.IG(.5,\tau^2/2)$} & \cellcolor{gray!6}{0.7} & \cellcolor{gray!6}{9312} & \cellcolor{gray!6}{14219} & \cellcolor{gray!6}{1.000} & \cellcolor{gray!6}{1.000} & \cellcolor{gray!6}{0.00} & \cellcolor{gray!6}{100} & \cellcolor{gray!6}{0}\\
\cellcolor{gray!6}{$6.HT(1,\tau)$} & \cellcolor{gray!6}{0.7} & \cellcolor{gray!6}{4} & \cellcolor{gray!6}{13705} & \cellcolor{gray!6}{1.006} & \cellcolor{gray!6}{1.420} & \cellcolor{gray!6}{42.14} & \cellcolor{gray!6}{46} & \cellcolor{gray!6}{2507}\\
$7.IG(2,2\tau^2)$ & 0.7 & 9488 & 14578 & 1.000 & 1.000 & 0.00 & 100 & 0\\
$8.HT(4,\tau)$ & 0.7 & 431 & 13673 & 1.000 & 1.013 & 10.30 & 50 & 466\\
\cellcolor{gray!6}{$9.IG(5,5\tau^2)$} & \cellcolor{gray!6}{0.7} & \cellcolor{gray!6}{10339} & \cellcolor{gray!6}{14484} & \cellcolor{gray!6}{1.000} & \cellcolor{gray!6}{1.001} & \cellcolor{gray!6}{0.00} & \cellcolor{gray!6}{100} & \cellcolor{gray!6}{0}\\
\cellcolor{gray!6}{$10.HT(10,\tau)$} & \cellcolor{gray!6}{0.7} & \cellcolor{gray!6}{4} & \cellcolor{gray!6}{13369} & \cellcolor{gray!6}{1.004} & \cellcolor{gray!6}{1.396} & \cellcolor{gray!6}{31.86} & \cellcolor{gray!6}{49} & \cellcolor{gray!6}{2512}\\
$11.IG(2,2(c\tau)^2)$ & 0.7 & 9102 & 14379 & 1.000 & 1.000 & 0.00 & 100 & 0\\
$12.HT(4,c\tau)$ & 0.7 & 426 & 13712 & 1.000 & 1.012 & 3.14 & 54 & 72\\
\cellcolor{gray!6}{$13.IG(2,2(c/\tau)^2)$} & \cellcolor{gray!6}{0.7} & \cellcolor{gray!6}{9666} & \cellcolor{gray!6}{14522} & \cellcolor{gray!6}{1.000} & \cellcolor{gray!6}{1.000} & \cellcolor{gray!6}{0.00} & \cellcolor{gray!6}{100} & \cellcolor{gray!6}{0}\\
\cellcolor{gray!6}{$14.HT(4,(\tau/c)^2)$} & \cellcolor{gray!6}{0.7} & \cellcolor{gray!6}{188} & \cellcolor{gray!6}{13513} & \cellcolor{gray!6}{1.000} & \cellcolor{gray!6}{1.024} & \cellcolor{gray!6}{10.40} & \cellcolor{gray!6}{47} & \cellcolor{gray!6}{252}\\\hline
\cellcolor{gray!6}{$1.IG(1,1)$} & \cellcolor{gray!6}{0.7} & \cellcolor{gray!6}{8409} & \cellcolor{gray!6}{12294} & \cellcolor{gray!6}{1} & \cellcolor{gray!6}{1.001} & \cellcolor{gray!6}{0} & \cellcolor{gray!6}{100} & \cellcolor{gray!6}{0}\\
\cellcolor{gray!6}{$2.IG(.001,.001)$} & \cellcolor{gray!6}{0.7} & \cellcolor{gray!6}{7600} & \cellcolor{gray!6}{12342} & \cellcolor{gray!6}{1} & \cellcolor{gray!6}{1.001} & \cellcolor{gray!6}{0} & \cellcolor{gray!6}{100} & \cellcolor{gray!6}{0}\\
$3.HT(1,1.2\tau)$ & 0.7 & 8369 & 12290 & 1 & 1.001 & 0 & 100 & 0\\
$4.HT(4,1)$ & 0.7 & 7770 & 12147 & 1 & 1.001 & 0 & 100 & 0\\
\cellcolor{gray!6}{$5.IG(.5,\tau^2/2)$} & \cellcolor{gray!6}{0.7} & \cellcolor{gray!6}{5961} & \cellcolor{gray!6}{11693} & \cellcolor{gray!6}{1} & \cellcolor{gray!6}{1.000} & \cellcolor{gray!6}{0} & \cellcolor{gray!6}{100} & \cellcolor{gray!6}{0}\\
\cellcolor{gray!6}{$6.HT(1,\tau)$} & \cellcolor{gray!6}{0.7} & \cellcolor{gray!6}{8521} & \cellcolor{gray!6}{12581} & \cellcolor{gray!6}{1} & \cellcolor{gray!6}{1.001} & \cellcolor{gray!6}{0} & \cellcolor{gray!6}{100} & \cellcolor{gray!6}{0}\\
$7.IG(2,2\tau^2)$ & 0.7 & 8442 & 11976 & 1 & 1.001 & 0 & 100 & 0\\
$8.HT(4,\tau)$ & 0.7 & 7736 & 11789 & 1 & 1.001 & 0 & 100 & 0\\
\cellcolor{gray!6}{$9.IG(5,5\tau^2)$} & \cellcolor{gray!6}{0.7} & \cellcolor{gray!6}{8657} & \cellcolor{gray!6}{11902} & \cellcolor{gray!6}{1} & \cellcolor{gray!6}{1.001} & \cellcolor{gray!6}{0} & \cellcolor{gray!6}{100} & \cellcolor{gray!6}{0}\\
\cellcolor{gray!6}{$10.HT(10,\tau)$} & \cellcolor{gray!6}{0.7} & \cellcolor{gray!6}{8022} & \cellcolor{gray!6}{12022} & \cellcolor{gray!6}{1} & \cellcolor{gray!6}{1.001} & \cellcolor{gray!6}{0} & \cellcolor{gray!6}{100} & \cellcolor{gray!6}{0}\\
$11.IG(2,2(c\tau)^2)$ & 0.7 & 7279 & 12277 & 1 & 1.000 & 0 & 100 & 0\\
$12.HT(4,c\tau)$ & 0.7 & 7889 & 12206 & 1 & 1.001 & 0 & 100 & 0\\
\cellcolor{gray!6}{$13.IG(2,2(c/\tau)^2)$} & \cellcolor{gray!6}{0.7} & \cellcolor{gray!6}{7164} & \cellcolor{gray!6}{11736} & \cellcolor{gray!6}{1} & \cellcolor{gray!6}{1.001} & \cellcolor{gray!6}{0} & \cellcolor{gray!6}{100} & \cellcolor{gray!6}{0}\\
\cellcolor{gray!6}{$14.HT(4,(\tau/c)^2)$} & \cellcolor{gray!6}{0.7} & \cellcolor{gray!6}{8501} & \cellcolor{gray!6}{11967} & \cellcolor{gray!6}{1} & \cellcolor{gray!6}{1.000} & \cellcolor{gray!6}{0} & \cellcolor{gray!6}{100} & \cellcolor{gray!6}{0}\\
\bottomrule
\end{tabular}
\caption{\label{tab:model3-diag-taur}Diagnostics for $\tau_r$ under Model 3, the multiple outcomes model, when $\tau_b = .04$ (top half) and $\tau_b = .16$ (bottom half).}
\end{table}

\begin{table}[!bh]
\centering
\fontsize{10}{12}\selectfont
\begin{tabular}[t]{lrrrrrrrr}
\toprule
Prior & \makecell[c]{True\\Value} & Mean & Median & Bias & \makecell[c]{Rel\\Bias} & RMSE & Cov & \makecell[c]{Interval\\Length}\\
\midrule
\cellcolor{gray!6}{$1.IG(1,1)$} & \cellcolor{gray!6}{0.7} & \cellcolor{gray!6}{0.700} & \cellcolor{gray!6}{0.700} & \cellcolor{gray!6}{0.000} & \cellcolor{gray!6}{0.001} & \cellcolor{gray!6}{0.019} & \cellcolor{gray!6}{0.97} & \cellcolor{gray!6}{0.084}\\
\cellcolor{gray!6}{$2.IG(.001,.001)$} & \cellcolor{gray!6}{0.7} & \cellcolor{gray!6}{0.699} & \cellcolor{gray!6}{0.699} & \cellcolor{gray!6}{-0.001} & \cellcolor{gray!6}{-0.001} & \cellcolor{gray!6}{0.019} & \cellcolor{gray!6}{0.96} & \cellcolor{gray!6}{0.084}\\
$3.HT(1,1.2\tau)$ & 0.7 & 0.699 & 0.699 & -0.001 & -0.001 & 0.019 & 0.96 & 0.084\\
$4.HT(4,1)$ & 0.7 & 0.699 & 0.699 & -0.001 & -0.001 & 0.019 & 0.96 & 0.084\\
\cellcolor{gray!6}{$5.IG(.5,\tau^2/2)$} & \cellcolor{gray!6}{0.7} & \cellcolor{gray!6}{0.699} & \cellcolor{gray!6}{0.699} & \cellcolor{gray!6}{-0.001} & \cellcolor{gray!6}{-0.001} & \cellcolor{gray!6}{0.019} & \cellcolor{gray!6}{0.97} & \cellcolor{gray!6}{0.084}\\
\cellcolor{gray!6}{$6.HT(1,\tau)$} & \cellcolor{gray!6}{0.7} & \cellcolor{gray!6}{0.699} & \cellcolor{gray!6}{0.699} & \cellcolor{gray!6}{-0.001} & \cellcolor{gray!6}{-0.002} & \cellcolor{gray!6}{0.019} & \cellcolor{gray!6}{0.96} & \cellcolor{gray!6}{0.084}\\
$7.IG(2,2\tau^2)$ & 0.7 & 0.699 & 0.699 & -0.001 & -0.001 & 0.019 & 0.97 & 0.084\\
$8.HT(4,\tau)$ & 0.7 & 0.699 & 0.699 & -0.001 & -0.001 & 0.019 & 0.96 & 0.084\\
\cellcolor{gray!6}{$9.IG(5,5\tau^2)$} & \cellcolor{gray!6}{0.7} & \cellcolor{gray!6}{0.699} & \cellcolor{gray!6}{0.699} & \cellcolor{gray!6}{-0.001} & \cellcolor{gray!6}{-0.001} & \cellcolor{gray!6}{0.019} & \cellcolor{gray!6}{0.96} & \cellcolor{gray!6}{0.084}\\
\cellcolor{gray!6}{$10.HT(10,\tau)$} & \cellcolor{gray!6}{0.7} & \cellcolor{gray!6}{0.699} & \cellcolor{gray!6}{0.699} & \cellcolor{gray!6}{-0.001} & \cellcolor{gray!6}{-0.001} & \cellcolor{gray!6}{0.020} & \cellcolor{gray!6}{0.96} & \cellcolor{gray!6}{0.084}\\
$11.IG(2,2(c\tau)^2)$ & 0.7 & 0.702 & 0.702 & 0.002 & 0.003 & 0.019 & 0.95 & 0.084\\
$12.HT(4,c\tau)$ & 0.7 & 0.699 & 0.699 & -0.001 & -0.001 & 0.019 & 0.96 & 0.084\\
\cellcolor{gray!6}{$13.IG(2,2(c/\tau)^2)$} & \cellcolor{gray!6}{0.7} & \cellcolor{gray!6}{0.698} & \cellcolor{gray!6}{0.697} & \cellcolor{gray!6}{-0.002} & \cellcolor{gray!6}{-0.003} & \cellcolor{gray!6}{0.019} & \cellcolor{gray!6}{0.95} & \cellcolor{gray!6}{0.084}\\
\cellcolor{gray!6}{$14.HT(4,(\tau/c)^2)$} & \cellcolor{gray!6}{0.7} & \cellcolor{gray!6}{0.698} & \cellcolor{gray!6}{0.698} & \cellcolor{gray!6}{-0.002} & \cellcolor{gray!6}{-0.002} & \cellcolor{gray!6}{0.019} & \cellcolor{gray!6}{0.97} & \cellcolor{gray!6}{0.084}\\\hline
\cellcolor{gray!6}{$1.IG(1,1)$} & \cellcolor{gray!6}{0.7} & \cellcolor{gray!6}{0.702} & \cellcolor{gray!6}{0.701} & \cellcolor{gray!6}{0.002} & \cellcolor{gray!6}{0.003} & \cellcolor{gray!6}{0.02} & \cellcolor{gray!6}{0.97} & \cellcolor{gray!6}{0.084}\\
\cellcolor{gray!6}{$2.IG(.001,.001)$} & \cellcolor{gray!6}{0.7} & \cellcolor{gray!6}{0.700} & \cellcolor{gray!6}{0.700} & \cellcolor{gray!6}{0.000} & \cellcolor{gray!6}{0.001} & \cellcolor{gray!6}{0.02} & \cellcolor{gray!6}{0.96} & \cellcolor{gray!6}{0.085}\\
$3.HT(1,1.2\tau)$ & 0.7 & 0.700 & 0.700 & 0.000 & 0.001 & 0.02 & 0.96 & 0.084\\
$4.HT(4,1)$ & 0.7 & 0.701 & 0.700 & 0.001 & 0.001 & 0.02 & 0.96 & 0.084\\
\cellcolor{gray!6}{$5.IG(.5,\tau^2/2)$} & \cellcolor{gray!6}{0.7} & \cellcolor{gray!6}{0.700} & \cellcolor{gray!6}{0.700} & \cellcolor{gray!6}{0.000} & \cellcolor{gray!6}{0.001} & \cellcolor{gray!6}{0.02} & \cellcolor{gray!6}{0.96} & \cellcolor{gray!6}{0.084}\\
\cellcolor{gray!6}{$6.HT(1,\tau)$} & \cellcolor{gray!6}{0.7} & \cellcolor{gray!6}{0.700} & \cellcolor{gray!6}{0.700} & \cellcolor{gray!6}{0.000} & \cellcolor{gray!6}{0.001} & \cellcolor{gray!6}{0.02} & \cellcolor{gray!6}{0.96} & \cellcolor{gray!6}{0.085}\\
$7.IG(2,2\tau^2)$ & 0.7 & 0.700 & 0.700 & 0.000 & 0.001 & 0.02 & 0.96 & 0.084\\
$8.HT(4,\tau)$ & 0.7 & 0.700 & 0.700 & 0.000 & 0.001 & 0.02 & 0.96 & 0.085\\
\cellcolor{gray!6}{$9.IG(5,5\tau^2)$} & \cellcolor{gray!6}{0.7} & \cellcolor{gray!6}{0.700} & \cellcolor{gray!6}{0.700} & \cellcolor{gray!6}{0.000} & \cellcolor{gray!6}{0.001} & \cellcolor{gray!6}{0.02} & \cellcolor{gray!6}{0.96} & \cellcolor{gray!6}{0.084}\\
\cellcolor{gray!6}{$10.HT(10,\tau)$} & \cellcolor{gray!6}{0.7} & \cellcolor{gray!6}{0.700} & \cellcolor{gray!6}{0.700} & \cellcolor{gray!6}{0.000} & \cellcolor{gray!6}{0.001} & \cellcolor{gray!6}{0.02} & \cellcolor{gray!6}{0.96} & \cellcolor{gray!6}{0.085}\\
$11.IG(2,2(c\tau)^2)$ & 0.7 & 0.704 & 0.703 & 0.004 & 0.005 & 0.02 & 0.97 & 0.084\\
$12.HT(4,c\tau)$ & 0.7 & 0.701 & 0.700 & 0.001 & 0.001 & 0.02 & 0.97 & 0.085\\
\cellcolor{gray!6}{$13.IG(2,2(c/\tau)^2)$} & \cellcolor{gray!6}{0.7} & \cellcolor{gray!6}{0.699} & \cellcolor{gray!6}{0.699} & \cellcolor{gray!6}{-0.001} & \cellcolor{gray!6}{-0.002} & \cellcolor{gray!6}{0.02} & \cellcolor{gray!6}{0.96} & \cellcolor{gray!6}{0.084}\\
\cellcolor{gray!6}{$14.HT(4,(\tau/c)^2)$} & \cellcolor{gray!6}{0.7} & \cellcolor{gray!6}{0.700} & \cellcolor{gray!6}{0.699} & \cellcolor{gray!6}{0.000} & \cellcolor{gray!6}{0.000} & \cellcolor{gray!6}{0.02} & \cellcolor{gray!6}{0.96} & \cellcolor{gray!6}{0.084}\\
\bottomrule
\end{tabular}
\caption{\label{tab:model3-estimates-taur}Summary results for $\tau_r$ under Model 3, the multiple outcomes model, when $\tau_b=0.04$ (top half) and $\tau_b=0.16$ (bottom half).}
\end{table}


\end{document}